\documentclass[aps,floatfix,eqsecnum,showpacs,preprint,tightenlines]{revtex4}
\usepackage{epsfig}
\usepackage{bm}

\usepackage{amsmath}

\usepackage{color}

\def\e{\kern+.5ex\lower.42ex\hbox{$\scriptstyle \iota$}\kern-1.10ex e}
\def\registered{{\ooalign{\hfil\raise .00ex\hbox{\scriptsize R}\hfil\crcr\mathhexbox20D}}}

	\newcommand{\BA}[1]{\langle #1 \mid}

\newcommand{\KT}[1]{\mid #1 \rangle}

\begin{document}


\title{Pion absorption from the lowest atomic orbital in $^2$H, $^3$H and $^3$He}



\author{J. Golak}
\affiliation{M. Smoluchowski Institute of Physics, Jagiellonian University, PL-30348 Krak\'ow, Poland}
\author{V. Urbanevych}
\affiliation{M. Smoluchowski Institute of Physics, Jagiellonian University, PL-30348 Krak\'ow, Poland}
\author{R. Skibi{\'n}ski}
\affiliation{M. Smoluchowski Institute of Physics, Jagiellonian University, PL-30348 Krak\'ow, Poland}
\author{H. Wita{\l}a}
\affiliation{M. Smoluchowski Institute of Physics, Jagiellonian University, PL-30348 Krak\'ow, Poland}
\author{K. Topolnicki}
\affiliation{M. Smoluchowski Institute of Physics, Jagiellonian University, PL-30348 Krak\'ow, Poland}
\author{V. Baru}
\affiliation{Ruhr-Universit\"at Bochum, Fakult\"at f\"ur Physik und Astronomie, Institut f\"ur Theoretische Physik II, D-44780 Bochum, Germany}
\author{A. A. Filin}
\affiliation{Ruhr-Universit\"at Bochum, Fakult\"at f\"ur Physik und Astronomie, Institut f\"ur Theoretische Physik II, D-44780 Bochum, Germany}
\author{E. Epelbaum}
\affiliation{Ruhr-Universit\"at Bochum, Fakult\"at f\"ur Physik und Astronomie, Institut f\"ur Theoretische Physik II, D-44780 Bochum, Germany}
\author{H. Kamada}
\affiliation{Department of Physics, Faculty of Engineering,
Kyushu Institute of Technology, Kitakyushu 804-8550, Japan}
\author{A. Nogga}
\affiliation{Institut f\"ur Kernphysik (IKP-3), 
Institute for Advanced Simulation (IAS-4),
Center for Advanced Simulation and Analytics (CASA), 
and J\"ulich Center for Hadron Physics (JCHP), Forschungszentrum J\"ulich,
D-52425 J\"ulich, Germany}


\date{\today}

\begin{abstract}
The
$\pi^- + {^2{\rm H}} \rightarrow n + n$, 
$\pi^- + {^3{\rm H}} \rightarrow n + n + n$,
$\pi^- + {^3{\rm He}} \rightarrow n + d$ 
and 
$\pi^- + {^3{\rm He}} \rightarrow p + n + n$
capture reactions from the lowest atomic orbitals
are studied under full inclusion of final state interactions.
Our results are obtained with the single-nucleon and two-nucleon 
transition operators derived at leading order in chiral effective field theory. 
The initial and final three-nucleon states are calculated 
with the chiral nucleon-nucleon SMS potential up to N$^4$LO$^+$ augmented by the consistently regularized  chiral N$^2$LO
three-nucleon potential. We found that absorption rates depend strongly
on the nuclear pion absorption operator used, and its two-body parts change the rates by a few orders of magnitude.
The final state interactions between nucleons generated by the two-nucleon forces are also important, while the three-nucleon interaction plays a visible role only in the $\pi^- + {^3{\rm He}} \rightarrow n + d$ reaction. 
Our absorption rate for the 
$\pi^- + {^2{\rm H}} \rightarrow n + n$ process is in good agreement with 
 the experimental data from the hadronic ground-state broadening in pionic 
	deuterium. The capture rates  on  $^3{\rm He}$ are also generally consistent with the  spectroscopic data within error bars, though 
	our central values are found to be systematically below the data.
We show that for the three-body breakup processes the dominant contributions to the absorption rates arise from the quasi-free scattering and final-state interaction kinematical configurations.
\end{abstract}


\maketitle


\section{Introduction}
\label{section1}

Pion production in nucleon-nucleon collisions and the
  related pion absorption reactions involving nuclei have been
  extensively studied both experimentally and theoretically. 
  In the early 1990s, the 
precise experimental data for the total cross section of neutral pion
production in proton-proton collisions in the threshold kinematics measured
at IUCF \cite{Meyer:1990yf} revealed a serious disagreement with
the theoretical calculations by almost an order of magnitude
\cite{Miller:1991ndb,Koltun:1965yk}.  
While it was quickly realized that the production mechanism
involving the Weinberg-Tomozawa vertex 
should play an important role for  charged pion
production \cite{Koltun:1965yk},  
this  phenomenological model still failed to describe the data by a factor of 2.
Various phenomenological attempts to account for the missing physics were made 
including the heavy meson exchanges, the off-shell $\pi N$ effects,  excitations of baryon resonances etc. -- 
 see  Ref.~\cite{Hanhart:2003pg} for a review and references therein.
These unexpected findings also
stimulated extensive theoretical research in the
framework of chiral effective field theory (EFT), see \cite{Baru:2013zpa} for a
review.

The studies of the $NN \to NN \pi$ reactions have revealed a
number of interesting  aspects. The first tree-level EFT calculations, carried out up to next-to-next-to leading order (N$^2$LO) 
in the original Weinberg counting \cite{Weinberg:1992yk}, did not actually show any significant improvements 
\cite{Cohen:1995cc,Park:1995ku,Sato:1997ps,daRocha:1999dm,Hanhart:1997jd}. 
Moreover, the N$^2$LO corrections at one-loop order~\cite{Dmitrasinovic:1999cu,Ando:2000ema} were found to be even larger  than the NLO ones, 
calling  the convergence of the Weinberg EFT into question \cite{Bernard:1998sz}.
On the other hand, one soon realized the important role
played by the momentum scale $p \sim \sqrt{M_\pi M}$ associated with
real pion production
\cite{Cohen:1995cc,daRocha:1999dm,Hanhart:2000gp,Hanhart:2002bu}.
Here, $M_\pi$ and $M$ refer to the pion and
nucleon masses, respectively. In particular, the order of magnitude differences between various one-loop diagrams of Ref.~\cite{Dmitrasinovic:1999cu} could 
 be naturally understood if this new scale is included in the power counting~\cite{Hanhart:2003pg}. 
Thus, the appearance of this scale requires the
appropriate modification of the chiral power counting as compared to
the standard framework used to describe few-nucleon reactions below
pion-production threshold, see Refs.~\cite{Epelbaum:2008ga,
  Machleidt:2011zz,Epelbaum:2019kcf} for review articles. 
 The use of this new counting scheme, now known as the momentum counting scheme (MCS), 
  led to a very good understanding of   the threshold charge pion production data   
  already at leading order in the MCS (LO-MCS)~\cite{Lensky:2005jc}.
Later on, the calculation was pushed forward to the N$^2$LO-MCS~\cite{Filin:2012za,Filin:2013uma,Baru:2016kru}. 
  
  Another interesting aspect of  P-wave pion production in $NN$ collisions is the
fact that the leading short-range mechanism is the same as the one in
the dominant three-nucleon force \cite{Epelbaum:2002vt}, the axial-vector current
operator \cite{Park:2002yp,Gazit:2008ma,Gardestig:2006hj,Krebs:2016rqz} 
as well as  in the reactions $\gamma  \, {\rm ^2H}\to\pi 
NN$~\cite{Lensky:2005hb,Lensky:2007zc} and $\pi\, {\rm ^2H}\to \gamma NN$~\cite{Gardestig:2006hj}.
 Thus, the reaction $NN \to NN \pi$ can, at least in
principle, be used to determine the corresponding low-energy
constant $c_D$ and to test consistency of chiral EFT. More recently,
pion production in two-nucleon collisions gained considerable interest
in connection with charge-symmetry breaking. Specifically, the
experimental measurement of the charge-symmetry breaking forward-backward asymmetry in the
$pn \to d \pi^0$ reaction  \cite{Opper:2003sb} has been argued  to provide
access to the strong-interaction proton-neutron mass difference
\cite{vanKolck:2000ip}.  This quantity  was extracted in  Ref.~\cite{Filin:2009yh} based on the    
$A_{fb}(pn\to d\pi^0)$ data, see also Ref.~\cite{Bolton:2009rq} for a related study.     
The cross section in the reaction $dd \to {\rm
  ^4He} \, \pi^0$ measured at IUCF and later also by the WASA-at-COSY collaboration \cite{WASA-at-COSY:2014pry,WASA-at-COSY:2017ybz}
is yet another very clean probe of charge symmetry
breaking, but it still lacks a reliable theoretical analysis due to the
required complicated treatement of the four-nucleon continuum, see
however \cite{Nogga:2006cp} for a first step along this line. 

Experimental data for near-threshold  pion production in nucleon-deuteron collisions are
also available, see e.g.~\cite{Cameron:1981dm,Bilger:2002aw,Dymov:2016uyu} as well as 
spectroscopic data on $\pi \,^3$He atoms \cite{Schwanner:1984sg,Branchings3He,McCarthy:1975zzb,Gotta1995}, but theoretical efforts are
very scarce as compared with the two-nucleon sector, see 
e.g.~\cite{Schneider:2002sd,Canton:2004zq}. In particular, we are not aware
of any calculations in the framework of chiral EFT. 

In this paper we focus on the inverse of the pion production processes
mentioned above. Specifically, we perform an exploratory study of
stopped $\pi^-$ absorption out of the lowest orbitals of
$^2$H, $^3$H and $^3$He
pionic atoms. Pion absorption is an important subprocess for the 
pion scattering reactions that gained a considerable interest
in connection with the determination of the corresponding scattering
lengths. It does not only govern the imaginary part of the scattering
length, but also contributes to its real part via the so-called
dispersive corrections. Chiral EFT calculations of the pion-deuteron
scattering length and the extraction of the pion-nucleon scattering
lengths can be found in Refs.~\cite{Lensky:2006wd,Baru:2011bw,Baru:2010xn}. For related studies of 
pion production in proton-proton collisions and 
the $\pi
 - ^3$He and $\pi - ^4$He scattering lengths 
 see Refs.~\cite{Baru:2002cg,Liebig:2010ki}.  

Results presented here are obtained within the formalism of the Faddeev equations, which is 
one of standard approaches in three-nucleon studies. It was first applied to 3N bound states (3NBS) and nucleon-deuteron scattering, see~\cite{Glocklephysrep} and references therein. Over time, this formalism has been extended to other processes, like electron-3NBS scattering~\cite{physrep} and 3NBS photodisintegration~\cite{SkibinskiEPJA2005} and further to various electroweak 
processes including  muon capture ~\cite{PRC90.024001,PRC94.034002,Skibinski2016,Urbanevych2021},
neutrino scattering~\cite{neutrino2018,neutrino2019} and pion radiative capture~\cite{PRC98.054001}.
An important advantage of this formalism is its flexibility. It enables us to use various models of nuclear forces or currents, and to easily identify effects of various dynamical ingredients, like final state interactions or three-nucleon forces.

The paper is organized in the following way.
In Sec.~\ref{section2} we introduce the single-nucleon and two-nucleon
pion absorption operators, which we treat in momentum space.
In the following section we show results 
for the $\pi^- + {^2{\rm H}} \rightarrow n + n$ and
$\pi^- + {^3{\rm He}} \rightarrow n + d$ two-body reactions.
Our results for the three-body processes,
$\pi^- + {^3{\rm He}} \rightarrow p + n + n$ and
$\pi^- + {^3{\rm H}} \rightarrow n + n + n$,
are shown in Sec.~\ref{section4}, where we discuss in detail 
the way we calculate the total absorption rates 
and present various predictions for the differential absorption rates
calculated with different three-nucleon dynamics. 
In these calculations we
employ the chiral SMS nucleon-nucleon potentials up to {N$^4$LO$^+$}~\cite{SMS} 
and the N$^2$LO three-nucleon forces \cite{Maris2021}. 
Finally, Sec.~\ref{section5} contains some concluding remarks.

\section{The transition operator}
\label{section2}

In the negative pion absorption process we assume that the initial state 
$ \KT{i\, } $ consists of the atomic $K$-shell pion wave function 
$ \KT{ \phi \, } $ 
and the initial nucleus state with the three-momentum ${\bf P}_i$ 
and the spin projection $m_{i}$:
\begin{eqnarray}
\KT{i\, } = \KT{\phi \, } \, \KT{\Psi_i \, {\bf P}_i \, m_{i} \, } \, .
\label{i}
\end{eqnarray}
The final state, $\KT{f\, }$,  is  
the nuclear state with the total 
three-momentum ${\bf P}_f$ and the set of spin projections $m_{f}$:
\begin{eqnarray}
\KT{f\, } = \KT{\Psi_f \, {\bf P}_f \, m_{f} \, } \, .
\label{f}
\end{eqnarray}

The transition from the initial to final state is 
given in terms of the nuclear matrix element $N$ 
of the nuclear transition operator 
$\rho$ between the initial and final nuclear states:
\begin{eqnarray}
N = 
\BA{\Psi_f \, {\bf P}_f \, m_{f} \, } \, 
\rho
\, \KT{\Psi_i \, {\bf P}_i \, m_{i} \, } \, .
\label{nlambda}
\end{eqnarray}
The $\rho$ operator contains single-nucleon (SN), two-nucleon (2N) 
and, in principle also many-nucleon (such as three-nucleon (3N)) 
contributions 
but in the present paper we restrict ourselves 
to the SN and 2N parts. We use the fact that the 3N operators are suppressed compared to the SN and 2N ones by the power counting.

As already pointed out in the introduction, the pion
  production operator has been extensively studied within the
  so-called momentum counting scheme (MCS), which is an extension of  the
  standard chiral EFT power counting to account for the momentum scale  
$p \sim \sqrt{M_\pi   M }$. In particular, the leading-order (LO-MCS)
contribution has been worked out in Ref.~\cite{Lensky:2005jc}. The corrections up
to next-to-next-to-leading order in the MCS have been derived in
Refs.~\cite{Filin:2012za,Filin:2013uma} including the explicit
contributions of the $\Delta$(1232) resonance and implemented in Ref.~\cite{Baru:2016kru}, see also Ref.~\cite{Baru:2013zpa} for a
review. These corrections have been implemented in Ref.~\cite{Baru:2016kru} to study  the cross section of the reaction $p+p\to  {^2{\rm H}} +\pi^+$ in the threshold
kinematics. 
While the results of Refs.~\cite{Lensky:2005jc,Baru:2016kru} showed good agreement with data,  they were obtained using 
modern phenomenological NN potentials only. On the other hand, it was argued in Ref.~\cite{Baru:2016kru} that 
the available at that time chiral interactions, regularized in a coordinate-space,   
 were generated with a cutoff, which tends to remove a part of the intermediate-range physics relevant for  pion production in NN collisions.

\begin{figure}
\includegraphics[width=9cm]{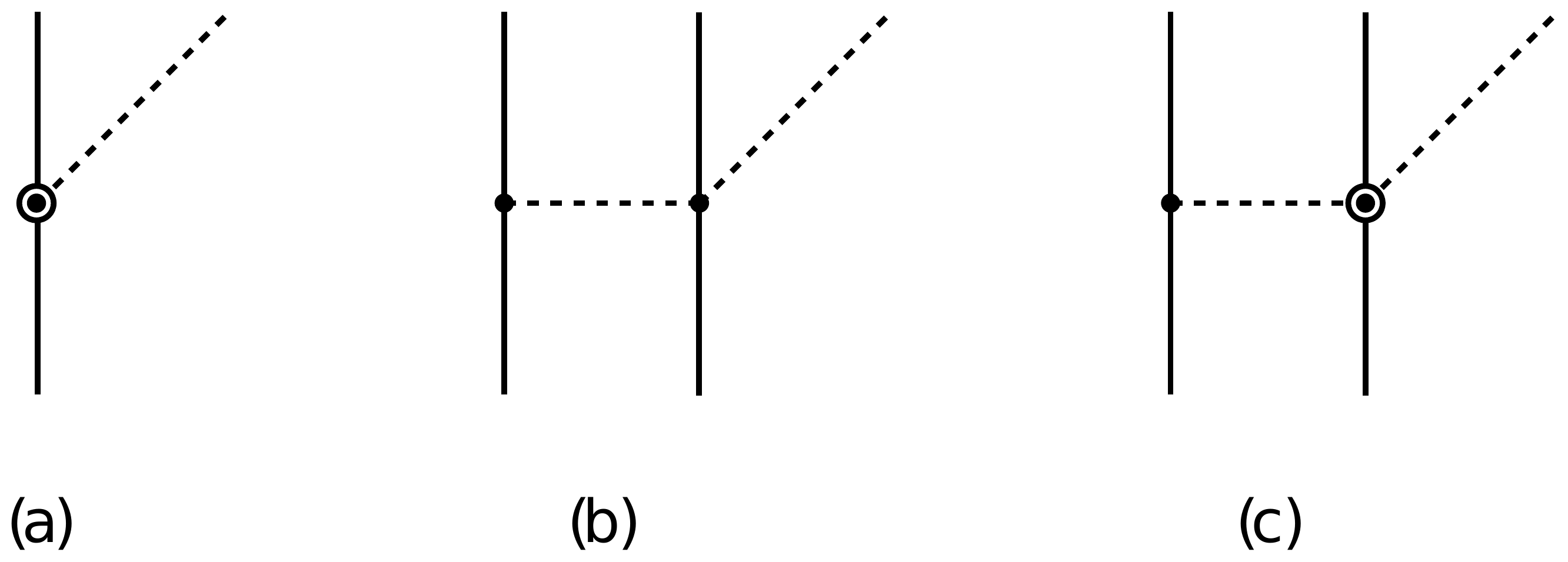}
\caption{Diagrams contributing to the pion production operator at LO
  in the MCS. Solid and dashed lines denote nucleons and pions,
  respectively. Solid dots are the pion-nucleon vertices from the
  lowest-order effective chiral Lagrangian, while circled dots denote
  the corresponding relativistic $1/M$-corrections.   
\label{Fig:Graphs}}
\end{figure}

 In this exploratory study, we would like to update the previous analysis by using the modern SMS chiral NN potentials \cite{SMS}, which are 
 softer than the coordinate-space regularized potentials and 
 therefore expected to be more suitable for studying the pion production process. 
 Furthermore, we will
use the calculation of  the $\pi^- + {^2{\rm H}} \rightarrow n + n$ channel,  which is directly  related to  $p+p\to  {^2{\rm H}}+ \pi^+$ via detailed balance, as benchmark 
to  present the first EFT-based predictions for the pion absorption rate on ${^3{\rm He}}$ and ${^3{\rm H}}$.
For these purposes, we limit ourselves
to the LO-MCS
contributions to the transition operator, which emerge from the direct
diagram (a) and the two rescattering graphs (b) and (c) as shown  in
Fig.~\ref{Fig:Graphs}. 
Notice that the static LO pion-nucleon vertex
proportional to the nucleon axial vector coupling $g_A$ does not
contribute to the direct term of type (a) due to the threshold
kinematics. We further emphasize that in addition to the rescattering
diagram (b) with the Weinberg-Tomozawa (WT) $\pi \pi NN$ vertex, one also has
to take into account diagram (c), which involves a $1/M$-correction to
the WT vertex. The appearance of both graphs at LO is the well-known
peculiar feature of the kinematics involved in the pion production
reaction. The explicit expressions for the SN (direct) and 2N parts of
the production operator are well known and will be specified below.
Notice further that we do not employ any regulator for the 2N
transition operator in this study. We also stress that even
though some of the numerical results discussed below are based on the nuclear forces
and the pion production operator derived in chiral EFT, our
approach is to be regarded as a hybrid one. This
is because the employed nuclear potentials are derived assuming the
kinematics with the nucleon momenta of the order of $\sim M_\pi$, and
they should, strictly speaking, be applied only below pion production
threshold.    

The momentum-space matrix element of the SN pion absorption operator $\rho(1)$  \cite{Bernard:1995dp}, for nucleon 1,
which depends on the nucleon incoming 
(${\bf p}$) 
and outgoing 
($ {\bf p}^{\, \prime}$) 
momentum,
is given by:
\begin{eqnarray}
\BA{{\bf p}^{\, \prime}} 
{\rho}(1) 
\KT{{\bf p}} = 
- \frac{g_A M_\pi}{\sqrt{2} F_\pi } \,
	\frac{ \left( {\bf p}^{\, \prime} +  {\bf p} \right) \cdot {\bm \sigma}_1 } { 2 M } \, 
	({\bm \tau}_{1})_- \, ,
\label{rho1}
\end{eqnarray}
where $g_A$=~1.29, $F_\pi$=~92.4 MeV, and $M_\pi$=~139.57 MeV/c$^2$ refer to the nucleon axial vector coupling,
pion decay constant, and negative pion mass, respectively.
$ \rho (1)$ is still an operator in the spin and 
isospin spaces, with ${\bm \sigma}_1$ (${\bm \tau}_1 $) being 
the Pauli spin (isospin) operator for nucleon~1 and 
the isospin lowering operator $({\bm \tau}_{1})_- \equiv(({\bm \tau}_1)_x -{\rm i} ({\bm \tau}_1)_y)/2$. 
The effects stemming from the small difference between the proton mass $M_p$ 
and neutron mass $M_n$ are beyond the accuracy level of our LO absorption operator and we use the average ``nucleon mass'', 
$M \equiv \frac12 \left( M_p + M_n \, \right) $.

We represent 2N and 3N states using partial wave decomposition and remind the reader of
its basic ingredients.
The 2N momentum space partial wave states,
$ \mid p \alpha_2 \, \rangle \equiv \mid  p (l s ) j m_j ; t m_t \, \rangle$,
carry information 
about the magnitude of the relative momentum ($p$),
the relative angular momentum ($l$), spin ($s$) and total angular
momentum ($j$) with the corresponding projection ($m_j$).
This set of quantum numbers is augmented by the 2N 
isospin ($t$) and its projection ($m_t$).

The corresponding ``$(2,3)1 $'' 3N states 
$ \mid p q \alpha \, \rangle \equiv \mid  p (l s ) j \, q (\lambda \frac12 )I 
\, (j I)J m_J ; T m_T \, \rangle $
are built on top of the subsystem $(2,3)$ quantum numbers 
and contain additionally information about 
nucleon 1: the magnitude of its relative momentum with 
respect to the c.m. of the $(2,3)$ subsystem ($q$),
relative orbital angular momentum ($\lambda$), total spin of nucleon 1 ($I$) 
and, finally, about the total 3N angular
momentum ($J$) with the projection ($m_J$) resulting from coupling 
of the 2-3 subsystem $j$ and spectator $I$ angular momenta.
The total 3N isospin state, 
$ \mid (t \frac12 ) T m_T \, \rangle$
is included analogously~\cite{physrep}.
Note that such full 3N partial-wave states are
already antisymmetrized in the $(2,3)$ subsystem.

In the case of the single-nucleon pion absorption operator applied to the initial 
deuteron state with the spin projection $m_d$ and total momentum ${\bf P}_i$ 
we can use a general formula~\cite{PRC90.024001} 
\begin{eqnarray}  
&&\langle p (l s) j m_j \, ; t m_t \, {\bf P}_f \mid \rho(1) \mid \phi_d \, m_d \, {\bf P}_i \, \rangle \ = \
\delta_{ t , 1 } \, 
\delta_{ m_t , -1 } \, 
\Big\langle  1\; {-1} \mid ({\bm \tau}_{1})_- \mid 0  0 \, \Big\rangle 
\nonumber \\
&& \times
\sum\limits_{m_l}  
	c \left( l , s , j ;  m_l,  m_j - m_l, m_j \,  \right) \,
\sum\limits_{l_d = 0 , 2}  \,
\sum\limits_{m_{l_d}}  
c \left( l_d , 1 , 1 ;  m_{l_d},  m_d - m_{l_d} , m_d \,  \right) \,
\nonumber \\
&& \times
\sum\limits_{m_1}  
c \left( \frac12 , \frac12 , s  ;  m_1,  m_j - m_l - m_1  , m_j - m_l  \,  \right) \,
\nonumber \\
&& \times
\sum\limits_{\mu_1}  
c \left( \frac12 , \frac12 , 1  ;  \mu_1,  m_d - m_{l_d} - \mu_1  , m_d - m_{l_d}   \,  \right) \,
\nonumber \\
&& \times \,
\delta_{ m_j - m_l - m_1  , m_d - m_{l_d} - \mu_1 \, } \,
\nonumber \\
&& \times
\int d \hat{\bf p} \,
Y^*_{l \, m_l } \left( \hat{\bf p} \right) \,
Y_{l_d \, m_{l_d} } \left( \widehat{{\bf p} - \frac12 {\bf Q} \, } \right) \,
\varphi_{l_d} \left( \mid {\bf p } - \frac12 {\bf Q} \mid \, \right) \,
\nonumber \\
&& \times \,
\Big\langle  \frac12 m_1 \mid \Big\langle \, {\bf p} + \frac12{\bf P}_f  
\mid \rho (1) \mid  {\bf p} - \frac12{\bf P}_f + {\bf P}_i \, \big\rangle 
\mid \frac12  \mu_1 \,  \Big\rangle \, ,
\label{pwdj1d}
\end{eqnarray}  
where $ {\bf Q} \equiv {\bf P}_f - {\bf P}_i $,
$ c \left( j_1 , j_2, j ; m_1 , m_2, m \, \right) $
are Clebsch-Gordan coefficients
and the internal deuteron state contains two components
\begin{eqnarray}  
\mid \phi_d \, m_d \, \rangle  =
\sum\limits_{l_d= 0, 2} \,
\int dp p^2  \, \mid p ( l_d 1 ) 1 m_d \, ;  0 0  \, \rangle \, \varphi_{l_d} \left( p \right) \, .
\label{pwdj1d2}
\end{eqnarray} 
We could in principle utilize the so-called automatized PWD method \cite{apwd1,apwd2}:
prepare momentum dependent spin matrix elements 
\begin{eqnarray}  
\Big\langle  \frac12 m^{\, \prime}  \mid \Big\langle {\bf p}_1^{\ \prime} 
\mid \rho(1) \mid  
{\bf p}_1 \, \Big\rangle \mid \frac12 m \, \Big\rangle \,
\label{pwdj1d3}
\end{eqnarray} 
and compute the double integral over
\begin{eqnarray}  
\int d \hat{\bf p} \, \equiv \,
	\int\limits_0^{2 \pi} d \phi_p \, 
	\int\limits_0^{\pi} d \theta_p \sin \theta_p 
\label{pwdj1d4}
\end{eqnarray} 
numerically.
However, for the absorption of a stopped pion on a nuclear system at rest we can assume here and in the following that ${\bf P}_i =  {\bf P}_f = 0 $ and benefit from the simple form of $\rho(1)$, which
leads to the conservation of the 2N total angular momentum 
and to the change of the 2N parity. As a result, matrix elements 
(\ref{pwdj1d}) are non-zero for just ONE partial wave state with 
$l=1$, $s=1$, $j=1$, and $m_j= m_d$:
\begin{eqnarray}  
\langle p (1 1) 1 m_j ; 1 -1 \, {\bf P}_f=0 \mid \rho(1) \mid \phi_d \, m_d \, {\bf P}_i=0 \, \rangle 
\nonumber \\
 = \ \delta_{ m_j , m_d } \, 
	\frac{g_A M_\pi}{2 \sqrt{2} M F_\pi } \, p \, \frac{
 2 \varphi_{0} \left( p \right)  + 
	\sqrt{2} \varphi_{2} \left( p \right) 
		}{\sqrt{3} } \, .
\label{pwdj1ana}
\end{eqnarray}

A general formula is also easy to obtain for a SN operator 
applied to a 3N bound state $\mid \Psi \, m_b \, ; \frac12 m_{T_b} \, {\bf P}_i \, \rangle$ 
with the spin $J_b = \frac12$, spin projection $m_b$, isospin $T_b = \frac12$, isospin projection $m_{T_b}$
and the total momentum ${\bf P}_i$, represented in the 3N partial wave states ${\alpha}_b$:
\begin{eqnarray}  
&&\langle p q \alpha \, {\bf P}_f \mid \rho(1) \mid \Psi \, m_b \, ; \frac12 m_{T_b} \, {\bf P}_i \, \rangle = 
\nonumber \\
	&&
	\sum\limits_{{\alpha}_b}
\delta_{ l  , l_b} \, 
\delta_{ s , s_b} \, 
\delta_{ j , j_b} \, 
\delta_{ t , t_b} \, 
	\delta_{ m_T , m_{T_b}-1 } \, 
	\Big\langle  \left( t \frac12 \right) T \, m_T \mid ({\bm \tau}_{1})_- \mid 
 \left( t_b \frac12 \right) \frac12 \, m_{T_b} \, \Big\rangle 
\nonumber \\
&& \times \,
\sum\limits_{m_j}  
c \left( j,  I, J ; m_j , m_J - m_j , m_J \,  \right) \,
c \left( j_b,  I_b, \frac12 ; m_j , m_b - m_j , m_b \,  \right) \,
\nonumber \\
&& \times \,
\sum\limits_{m_\lambda}  
c \left( \lambda , \frac12 , I ; m_\lambda , m_J - m_j - m_\lambda , m_J - m_j \,  \right) \,
\nonumber \\
&& \times \, 
\sum\limits_{m_{\lambda_b}}  
c \left( \lambda_b , \frac12 , I_b ; m_{\lambda_b} , m_b - m_j  - m_{\lambda_b} , m_b - m_j \,  \right) \,
\nonumber \\
&& \times
\int d \hat{\bf q} \,
Y^*_{\lambda\, m_\lambda } \left( \hat{\bf q} \right) \,
Y_{\lambda_b \, m_{\lambda_b} } \left( \widehat{{\bf q} - \frac23 {\bf Q} \, } \right) \,
\phi_{{\alpha}_b} \left( p , \mid {\bf q} - \frac23 {\bf Q} \mid \, \right)
\nonumber \\
&& \times \, \Big\langle  \frac12 \, m_J - m_j - m_\lambda \mid \Big\langle \, {\bf q} + \frac13{\bf P}_f  
\mid \rho(1) \mid  {\bf q} - \frac23{\bf P}_f + {\bf P}_i \, \big\rangle 
\mid \frac12 \, m_b - m_j - m_{\lambda_b} \,  \Big\rangle \,
\label{pwdj1}
\end{eqnarray}  
where, as in the 2N space, $ {\bf Q} \equiv {\bf P}_f - {\bf P}_i $.
The initial 3N bound state is given as
\begin{eqnarray}  
\mid \Psi \, m_b \, ; \frac12 m_{T_b} \, \rangle  =
\sum\limits_{{\alpha}_b} \,
\int dp p^2 \int dq q^2 \, 
\Big| p q {\alpha}_b \, \Big\rangle \, \phi_{{\alpha}_b} \left( p , q \, \right) \, .
\label{pwdj13}
\end{eqnarray}  
For $\rho(1)$ the result given in Eq.~(\ref{pwdj1}) can be simplified, especially 
for ${\bf P}_i =  {\bf P}_f = 0 $, and written as 
\begin{eqnarray}  
\langle p q \alpha \, {\bf P}_f=0 \mid \rho(1) \mid \Psi \, m_b \, ; \frac12 m_{T_b} \, {\bf P}_i=0 \, \rangle = 
\nonumber \\
	\frac{g_A M_\pi \sqrt{6} }{M F_\pi } \, q \,
	\delta_{ m_T , m_{T_b}-1 } \, 
	\delta_{ J , \frac12 } \, 
	\delta_{ m_J , m_b } \,
	\sqrt{\left( 2 \lambda + 1 \right) \, } \,
	(-1)^{I+\frac12} \,
\nonumber \\
\times \,
	(-1)^t \,
	\left\{
		\begin{array}{ccc}
			1 & \frac12 & \frac12 \\
			t & T & \frac12 
		\end{array}
		\right\} \, 
	c\left( 1, \frac12, T ; -1, m_{T_b} , m_{T_b}-1 \, \right) \,
\nonumber \\
\times \,
	\sum\limits_{{\alpha}_b}
\delta_{ l  , l_b} \, 
\delta_{ s , s_b} \, 
\delta_{ j , j_b} \, 
\delta_{ t , t_b} \, 
\delta_{ I , I_b} \, 
\phi_{{\alpha}_b} \left( p , q \, \right) \, 
\nonumber \\
\times \,
	\sqrt{ \left( 2 \lambda_b + 1 \right) \, } \,
	c\left( \lambda, \lambda_b , 1 ; 0, 0, 0 \, \right) \,
	\left\{
		\begin{array}{ccc}
			\lambda & \lambda_b & 1 \\
			\frac12 & \frac12 & I 
		\end{array}
		\right\} \, ,
\label{pwdj14}
\end{eqnarray}  
where the change of the 3N parity follows from the property
of the Clebsch-Gordan coefficient 
$c\left( \lambda, \lambda_b , 1 ; 0, 0, 0 \, \right) $,
which is non-zero only for odd $\lambda + \lambda_b$.

The 2N part of $ \rho$ at LO, stemming from the rescattering diagrams (b) and (c) of Fig.~\ref{Fig:Graphs},
has the form \cite{Lensky:2005jc}
\begin{eqnarray}
\BA{
	{\bf p}_1^{\, \prime}\,
	{\bf p}_2^{\, \prime}
	} 
{\rho}(1,2) 
\KT{
	{\bf p}_1 \, 
	{\bf p}_2 \, 
	} = 
	\left(
	v( k_2 )  {\bf k}_2 \cdot {\bm \sigma}_2 \, - \, 
	v( k_1 )  {\bf k}_1 \cdot {\bm \sigma}_1 \,
	\right) \, \nonumber \\ \times \,
	\frac{i}{\sqrt{2}} \, 
	\left[ 
	    \left( {\bm \tau}_1 \times {\bm \tau}_2 \, \right)_x 
	    - i \left( {\bm \tau}_1 \times {\bm \tau}_2 \, \right)_y \,
	\right] \,,
\label{rho12}
\end{eqnarray}
where 
$ {\bf k}_1 = {\bf p}_1^{\, \prime} - {\bf p}_1 $,
$ {\bf k}_2 = {\bf p}_2^{\, \prime} - {\bf p}_2 $
and the formfactor $v(k)$ reads 
\begin{eqnarray}
v (k) = \frac 1{ \left( 2 \pi \, \right)^3 } \,
        \frac{g_A M_\pi}{4 F_\pi^3 } \,
	\frac1{M_\pi^2 + k^2 } \, .
\label{vk}
\end{eqnarray}
For the operator defined in (\ref{rho12}) it is sufficient to calculate the corresponding matrix elements without isospin
\begin{eqnarray}
	H\left( p' , {\bar \alpha}_2^{\, \prime} ; p , {\bar \alpha}_2  \, \right) \, 
	\equiv 
	\, \langle p' (l' s') j' m_{j'}  \mid \rho(1,2)^{\rm spin} \mid p (l s) j m_j \, \rangle \, ,
\label{spinpart}
\end{eqnarray}
where 
\begin{eqnarray}
\BA{
	{\bf p}_1^{\, \prime}\,
	{\bf p}_2^{\, \prime}
	} 
	{\rho}(1,2)^{\rm spin}  
\KT{
	{\bf p}_1 \, 
	{\bf p}_2 \, 
	} = 
	v( k_2 )  {\bf k}_2 \cdot {\bm \sigma}_2 \, - \, 
	v( k_1 )  {\bf k}_1 \cdot {\bm \sigma}_1 \, ,
\label{rho121}
\end{eqnarray}
and supplement them later with appropriate 2N or 3N isospin matrix elements~\cite{physrep}.
For ${\bf P}_i =  {\bf P}_f = 0 $ one gets simply 
\begin{eqnarray}
v( k_2 )  {\bf k}_2 \cdot {\bm \sigma}_2 \, - \, 
v( k_1 )  {\bf k}_1 \cdot {\bm \sigma}_1 \,
= \, - v( k_1 )  {\bf k}_1 \cdot \left( {\bm \sigma}_1 + {\bm \sigma}_2 \, \right) 
	\label{j12s}
\end{eqnarray}
and the ${\bf k}_1$ vector is just the difference between the final and initial 
relative momenta, ${\bf k}_1 = {\bf p}^{\, \prime} - {\bf p} $.
Equation (\ref{j12s}) constitutes a convenient starting point for analytical
evaluation of $ H\left( p' , {\bar \alpha}_2^{\, \prime} ; p , {\bar \alpha}_2  \, \right) $.
Standard steps \cite{Gloecklebook} using multiple re-couplings of 
angular momenta, properties of the spherical harmonics and Clebsch-Gordan 
coefficients lead to 
\begin{eqnarray}
	\langle p' (l' s') j' m_{j'}  \mid \rho(2,3)^{\rm spin} \mid p (l s) j m_j \, \rangle = 
	\nonumber \\
\delta_{ j  , j' } \, 
	\delta_{ m_j , m_{j' } } \, 
\delta_{ s , 1} \, 
	\delta_{ s' , 1} \, 12 \pi \sqrt{2} \, 
	(-1)^j \,
	\left\{
		\begin{array}{ccc}
			l & l' & 1 \\
			1 & 1 & j 
		\end{array}
		\right\} \, 
	\nonumber \\
	\times \,
	\sum\limits_{a_1+a_2=1} (p')^{a_1} \, p^{a_2} \, (-1)^{a_2} 
	\sum\limits_{w} (2 w + 1) \, (-1)^w \, g_w (p',p ) \,
	\nonumber \\
	\times \,
	\left\{
		\begin{array}{ccc}
			l & l' & 1 \\
			a_1 & a_2 & w 
		\end{array}
		\right\} \, 
	c\left( w , a_1 , l' ; 0, 0, 0 \, \right) \,
	c\left( w , a_2 , l  ; 0, 0, 0 \, \right) \, ,
\end{eqnarray}
where 
\begin{eqnarray}
	g_w (p',p ) \, = \, \int\limits_{-1}^1 dx P_w (x) \, 
	v\left( \sqrt{ (p')^2 + p^2 - 2 p p' x \, } \right) \, ,
\end{eqnarray}
with $P_w(x)$ being Legendre polynomials.

Finally, for the reaction on the deuteron
we need only one isospin matrix element
\begin{eqnarray}
	\langle \left( \frac12 \frac12 \right) 1 -1 \mid
	\frac{i}{\sqrt{2}} \, 
	\left[ 
	    \left( {\bm \tau}_1 \times {\bm \tau}_2 \, \right)_x 
	    - i \left( {\bm \tau}_1 \times {\bm \tau}_2 \, \right)_y \,
	\right]  
	\mid 
	\left( \frac12 \frac12 \right) 0 0 \rangle \, = 2 \, ,
	\label{ISO12d}
\end{eqnarray}
while the missing 3N isospin matrix element to be used in the reactions 
with the 3N bound states is evaluated to 
\begin{eqnarray}
	\Big\langle  \left( t' \frac12 \right) T' \, m_{T'} \mid 
	\frac{i}{\sqrt{2}} \, 
	\left[ 
	    \left( {\bm \tau}_2 \times {\bm \tau}_3 \, \right)_x 
	    - i \left( {\bm \tau}_2 \times {\bm \tau}_3 \, \right)_y \,
	\right]  
	\mid 
 \left( t \frac12 \right) \frac12 \, m_T \, \Big\rangle  \nonumber \\
	=
	12\sqrt{3} \, \sqrt{ \left( 2 t + 1  \right) \, \left( 2 t' + 1  \right) \, } \,
	(-1)^{t + \frac32 + T'}  \,
	\left\{
		\begin{array}{ccc}
			1 & t & t' \\
			\frac12 & T' & \frac12 
		\end{array}
		\right\} \, 
		\nonumber \\
		\times \,
	\left\{
		\begin{array}{ccc}
			1 & 1 & 1 \\
			\frac12 & \frac12 & t \\
			\frac12 & \frac12 & t'
		\end{array}
		\right\} \, 
	c\left( 1, \frac12, T' ; -1, m_T , m_T-1 \, \right) \, 
	\label{ISO23}
\end{eqnarray}
and vanishes unless $t + t' = 1$.

\section{Results for the 
$\pi^- + {^2{\rm H}} \rightarrow n + n $ 
and
$\pi^- + {^3{\rm He}} \rightarrow n + d $ 
reactions}
\label{section3}
Recently, we investigated muon capture on $^2$H, $^3$He and $^3$H \cite{PRC90.024001,PRC94.034002}
as well as pion radiative capture in the same nuclei
\cite{PRC98.054001}. 
In the corresponding papers we described our momentum space framework and the way 
we calculate the initial and final nuclear states. We refer the reader especially to Appendices A 
and~B of Ref.~\cite{PRC98.054001} for details. Since our framework is consistently non-relativistic,
we also analyzed effects stemming from approximate non-relativistic treatment of kinematics.
The kinematics of the pion absorption processes studied in the present paper 
is in fact simpler than for pion radiative capture and can be treated 
as a special case of it with the final photon energy zero.
The initial state, including the K-shell pion, is the same.
Thus we can be very brief and provide only few numbers, for the reader's orientation.
Assuming the following values for the proton, neutron, deuteron, $^3$He
and negative pion masses, respectively,
$ M_p$ = 938.272 MeV,
$ M_n$ = 939.565 MeV,
$ M_d$ = 1875.613 MeV,
$ M_{3He}$ = 2808.392 MeV,
$M_{\pi^-}$ = 139.570 MeV,
our (nonrelativistic) results for the magnitudes
of the neutron momenta in the two-body pion absorption reactions are
$p_0$ = 357.534 MeV ($\pi^- + ^2{\rm H} \rightarrow n + n $) 
and
$q_0$ = 407.726 MeV 
($\pi^- + {^3{\rm He}} \rightarrow n + d $).
We neglected the small atomic binding energy of the pion.
The relativistic numbers are slightly (by approximately 2~\%) bigger.

Let us start with the $\pi^- + {^2{\rm H}} \rightarrow n + n $ reaction. 
The key ingredient of the absorption rate is here the nuclear matrix element
of the transition operator $\rho = \rho(1) + \rho(2) + \rho(1,2)$ 
between the initial deuteron state and the final two-neutron scattering state.
Introducing the spin magnetic quantum numbers $m_d$, $m_1$, $m_2$,
for the deuteron, neutron~1 and neutron~2, respectively, we write
\begin{eqnarray}
	 N_{nn} (m_1, m_2, m_d \, ) \ = \
^{(-)}\BA{ {\bf p}_0 \, m_{1} \, m_2 \ {\bf P}_f=0 } \,
\rho \, \KT{\phi_d \, m_d \ {\bf P}_i=0 \, } \, ,
\label{nnn1}
\end{eqnarray}
where $ \KT{  {\bf p}_0 \, m_{1} \, m_2 \ {\bf P}_f=0  }^{(-)}  $ denotes the 2N scattering state (see for example \cite{neutrino2018}).

Collecting all factors we arrive at the following 
expression for the total absorption rate in the
$\pi^- + {^2{\rm H}} \rightarrow n + n $ reaction \cite{bjodrell}:
 
\begin{eqnarray}
	&&\Gamma_{nn} = 
\frac{ \left( \alpha \, M^\prime_d \, \right)^3 \, c \, M_n \, p_0 }{ 2 M_{\pi^-} }
	  \int d {\bf\hat p}_0 \,
	  \frac13 \, 
	 \sum\limits_{m_1, m_2, m_d} 
	 \left| 
	 N_{nn} (m_1, m_2, m_d \, ) \, 
	 \right|^2  \, ,
\label{gnn1}
\end{eqnarray}  
where the phase-space factor, normalizations of the pion field and of the two-nucleon states as well as the $ \frac { \left(  M^\prime_d \alpha \, \right)^3 } {\pi  } $ factor stemming from the $K$-shell atomic wave function are taken into account with $ M^\prime_d  = \frac { M_d M_{\pi^-} } { M_d + M_{\pi^-} }$ and  $ \alpha \approx \frac 1{137} $ being the fine structure constant. 
The speed of light $c$ is used to convert the units of $\Gamma_{nn}$ from fm$^{-1}$ to inverse seconds.
We can further simplify (\ref{gnn1}), since for the unpolarized case there is no
dependence on the direction of the neutron momentum, ${\bf\hat p}_0$ and the integral over $d {\bf\hat p}_0$ yields $ 4 \pi$.

The results obtained with two different types of the two-nucleon potential,
with different treatment of the final two-neutron state
and the transition operator are collected in Table~\ref{tablenn}.
They show that the 2N contribution to the pion absorption operator changes the results (both PW and Full) obtained with the single-nucleon absorption operator by two-three orders of magnitude. Final-state interactions play an important role and their effects are especially strong in the calculations employing only the single-nucleon absorption operator. 
This conclusion is  consistent with the previous studies  of the time-reversed processes of meson production in NN collisions and  of the reaction $NN\to NN\pi $, in particular, where the related NN initial-state interactions are known to play an important role~\cite{Hanhart:2003pg,Baru:2013zpa}.  

The full results (SN+2N, Full) calculated with the N$^4$LO$^+$ chiral NN wave functions show  very good agreement with the experimental data.
Predictions computed for  the complete LO-MCS transition operator with $\Lambda$= 450 MeV  at different chiral orders for the NN wave functions show good convergence.
The spread of the complete results with the cutoff (measured by the standard deviation) is roughly two times smaller than for the corresponding (SN+2N, PW) calculations. Clearly, the results for the softest cutoff $\Lambda$= 400 MeV slightly deviate from the others. Pion absorption is sensitive to intermediate momentum components of the wave function, which might be not properly represented in the wave functions computed with small cutoff values. Therefore, one might consider rejecting results obtained with the smallest $\Lambda$ value.   
 On the other hand, the  cutoff dependence of the complete results stemming from the chiral NN wave functions at  N$^4$LO$^+$  is much weaker than the expected 
 LO-MCS theoretical uncertainty, even when the smallest cutoff  is included.  
 Indeed, using  the expansion parameter for the production operator, $\chi \sim p/ \Lambda_\chi \simeq 0.4$,
 where   $\Lambda_\chi \simeq 4 \pi f_{\pi} \simeq$1 GeV is the chiral symmetry breaking scale,  the   theoretical uncertainty   at the given order $\nu$ is expected to come from the first neglected chiral order $\nu +1$ unless the coefficient in front of this term vanishes. Therefore,  the   theoretical uncertainty for the amplitude at the LO-MCS can be estimated as $\sim \chi$, which translates to $\sim 2\chi \simeq 80\%$ for the absorption rate, where the amplitude enters squared.
  It should, however, be stressed that there are no pion production operators involving pions 
and nucleons at next-to-leading order in the MCS (NLO-MCS), since all NLO-MCS operators from 
the loop diagrams were shown to cancel exactly in Ref.~\cite{Lensky:2005jc}. Also, 
as discussed in Ref.~\cite{Baru:2016kru}, there is a significant cancellation of the 
one-loop and tree-level diagrams involving the $\Delta(1232)$ at NLO-MCS. Therefore, the 
actual truncation uncertainty 
for the absorption rate on the deuteron is at least a factor of 2 smaller.  We however keep this conservative uncertainty estimate
for the absorption rates on heavier nuclei, since the cancellation of various  $\Delta(1232)$ contributions 
 might not be operative in such systems.
  
  The results shown in Table~\ref{tablenn} can  be also compared with the state-of-the-art calculations of Refs.~\cite{Lensky:2005jc,Baru:2016kru}. 
  The cross section of the $p+p\to  {^2{\rm H}}+ \pi^+$ reaction in the center-of-mass system at threshold is conveniently parametrized as 
  \begin{equation}\label{XSthr}
  \sigma = \alpha \eta,  
  \end{equation}
  where $\eta$ is the outgoing pion momentum in the units of  the pion mass. Using detailed balance, the absorption rate $\Gamma_{nn}$ 
  of the $\pi^- + {^2{\rm H}} \rightarrow n + n $ channel can be  straightforwardly connected with the threshold parameter $\alpha$ in 
   Eq.~\eqref{XSthr}, see, e.g., Ref.~\cite{Strauch:2010vu} for details.  This yields  the experimental  
    value  from the hadronic ground-state broadening in pionic 
	deuterium to be $\alpha=251^{+5}_{-11} \mu b$. Our LO-MCS calculation with the N$^4$LO$^+$ chiral wave functions gives 
	 $\alpha=\{283, 246, 233, 238\} \mu b$ for $\Lambda=\{400, 450, 500, 550\}$ MeV, in order. These results agree very well with the corresponding results 
	 of Refs.~\cite{Lensky:2005jc,Baru:2016kru} obtained using modern phenomenological potentials. 	 
	 Therefore, we are now well prepared to make first EFT-based predictions for the absorption rates of more complicated pion capture reactions on 
	 $^3{\rm He}$ and $^3{\rm H}$.

\begin{table}
\caption{Absorption rates 
for the $\pi^- + {^2{\rm H}} \rightarrow n + n $ reaction
calculated with 
the chiral semilocal momentum-space regularized two-nucleon 
	force~\cite{SMS} at given chiral order (specified in column "chiral order in nuclear w.f.") and with 
	particular cutoff values (column "$\Lambda$")
	with the single-nucleon transition operator (SN) and 
	including two-nucleon contributions (SN+2N) at leading order in the MCS (column "chiral order in production").
Plane wave results (PW) and results obtained with the inclusion of 
	two-neutron rescattering (Full) are shown. Last column refers to the experimental value from the hadronic ground-state broadening in pionic
	deuterium~\cite{Strauch:2010rm,Strauch:2010vu} }
\begin{tabular}{||l|l|c|c|c|c|c|c||}
\hline\hline
\multicolumn{3}{||c|}{} & \multicolumn{5}{c||}{Absorption rate $\Gamma_{nn}$ in 10$^{15}$ s$^{-1}$}  
\\ \cline{4-8}
\multicolumn{3}{||c|}{} & \multicolumn{2}{c|}{SN} & \multicolumn{2}{c|}{SN+2N} & { Exp.}\\ \cline{3-6}
\hline
chiral order & chiral order & $\Lambda$ (MeV) & PW & Full & PW & Full &\\ 
in production&in nuclear w.f.&   &   &   &   &   &\\ 
\cline{1-7}
 LO-MCS &LO & 450  &     0.0593 &     0.0883 &      3.541 &      3.613  & \\ 
\cline{1-7}
LO-MCS & NLO & 450   &     0.0001 &     0.0135 &      2.221 &      2.059 &\\ 
\cline{1-7}
 LO-MCS &N$^2$LO & 450   &     0.0158 &     0.0039 &      1.827 &      1.433 &\\ 
\cline{1-7}
LO-MCS & N$^3$LO & 450   &     0.0155 &     0.0087 &      1.836 &      1.237& $1.306^{+0.026}_{-0.055}$ \\ 
\cline{1-7}
 LO-MCS &N$^4$LO & 450   &     0.0131 &     0.0091 &      1.850 &      1.243& \\ 
\cline{1-7}
 LO-MCS &{N$^4$LO$^+$} & 400   &     0.0028 &     0.0125 &      2.057 &      1.484 &\\ 
LO-MCS & {N$^4$LO$^+$} & 450   &     0.0142 &     0.0070 &      1.836 &      1.292 &\\ 
 LO-MCS &{N$^4$LO$^+$} & 500   &     0.0305 &     0.0032 &      1.644 &      1.224 &\\ 
 LO-MCS &{N$^4$LO$^+$} & 550   &     0.0460 &     0.0007 &      1.508 &      1.247 &\\
\hline\hline
\end{tabular}
\label{tablenn}
\end{table}

Let us now turn to the calculations of the absorption rate for the
$\pi^- + {^3{\rm He}} \rightarrow n + d $ reaction. Pion absorption in $^3$He was studied theoretically before in the early 1990s~\cite{Niskanen1991,Kiang1994}. 
In \cite{Niskanen1991} absorption of negative pions with energies between 42 MeV and 256 MeV on ${^1S_0}$ proton pairs in $^3$He was investigated in a model which included only partially the final-state interactions. 
The authors of \cite{Kiang1994} studied two-body pion absorption of stopped pions
including various short-range absorption mechanisms but neglecting the nuclear distortion
effects on the absorbed pions and the outgoing nucleons. Their work clearly showed the importance of the two-body terms in the pion absorption operator.  
In our formulation the crucial role is played by the matrix element
of the 3N transition operator $\rho_{3N}$ 
\begin{eqnarray}  
N_{nd} (m_n, m_d , m_{^3{\rm He}}  \, )  \, \equiv \, 
{{}^{(-)}\BA{\Psi_{nd}  \, 
	m_n \, m_d \,
	{\bf P}_f=0 
	}} \, 
	\rho_{3N}
\, \KT{\Psi_{^3{\rm He}} \, m_{^3{\rm He}} \, {\bf P}_i=0 \, 
	} 
\label{nnd}
\end{eqnarray} 
between the initial $^3$He and the final two-cluster 3N scattering state
with $\rho_{3N} = \rho(1) + \rho(2) + \rho(3) + 
\rho(2,3) + \rho(3,1) + \rho(1,2) + \rho(1,2,3) $ with the latter term neglected in this work, since its effects are expected to be beyond the accuracy level of our present calculations.
Our formula for the total absorption rate reads:

\begin{equation}
	\Gamma_{nd} = 
 {\cal R} \, \frac{ 16 \, \left( \alpha^3 \, M^\prime_{^3{\rm He}} \, \right)^3  \, c \, M q_0 }{ 9 M_{\pi^-}  }
	  \int d {\bf\hat q}_0 \,
	  \frac12 \, 
	 \sum\limits_{m_n, m_d, m_{^3{\rm He}}} 
	 \left| 
	 N_{nd} (m_n, m_d, m_{^3{\rm He}} \, ) \, 
	 \right|^2  \, ,  
\label{gnd}
\end{equation}  
where
$ M^\prime_{^3{\rm He}}  = \frac { M_{^3{\rm He}} M_{\pi^-} } { M_{^3{\rm He}} + M_{\pi^-} }$
is now the reduced mass of the $\pi^- - {^3{\rm He}}$ system.
We can use the same arguments as before to simplify the angular integrations. The
final state energy is expressed in terms of the neutron momentum ${\bf q}_0$
\begin{eqnarray}
M_\pi + M_{^3{\rm He}} \approx M_n + M_d + \frac34 \frac{ {\bf q}_0^{\ 2}} {M} \, ,
\label{eq0}
\end{eqnarray}  
where we neglect the deuteron binding energy in the kinetic energy and use the average nucleon mass $M$.
The factor ${\cal R} = 0.98 $ is due to the finite
volume of the $^3$He charge \cite{prc83.014002}.
(The corresponding factor in $\Gamma_{nn}$ is very close to $1$ \cite{prc83.014002} and has therefore not been included.)

Our results for $\Gamma_{nd}$ are shown in Table~\ref{tablend}.
We display there four different types of predictions obtained at N$^4$LO$^+$ for four values of the parameter $\Lambda$:  
	(1) symmetrized plane wave with the single-nucleon and two-nucleon 
	parts in the transition operator and three-nucleon force effects included 
	in the initial three-nucleon bound state (PWIAS--(SN+2N)--(2NF+3NF)),
	(2) calculation with the initial and final states calculated with the same 
	Hamiltonian comprising two- and three-nucleon forces but keeping only the single-nucleon contribution in the transition operator
	(Full--SN--(2NF+3NF)),
	(3) calculation with the initial and final states calculated with the same 
	Hamiltonian comprising only two-nucleon forces and including the single-nucleon
	and two-nucleon parts in the transition operator (Full--(SN+2N)--2NF), and
	(4) calculation with the initial and final states calculated with the
	two- and three-nucleon forces and the complete transition operator (Full--(SN+2N)--(2NF+3NF)). Thus by comparing calculations (2) and (4) we see the importance of the two-nucleon part of the transition operator. Also final state interactions play a significant role and reduce the rates approximately by a factor of three (calculations (1) vs. (4)). The three-nucleon force effects are clearly visible and grow with $\Lambda$ amounting to more than 30~\% for $\Lambda$= 550 MeV (calculations (3) vs. (4)). The dependence of our most complete calculations (4) on the cutoff parameter $\Lambda$ remains evident. This could be, however, expected since only LO MCS absorption operators are taken into account.
	Specifically, the cutoff dependence is expected to be significantly reduced when the  two NN$\rightarrow\textrm{NN}\pi$ contact terms at NNLO-MCS  will be taken into account.

\begin{table}
\caption{Absorption rates 
for the $\pi^- + {^3{\rm He}} \rightarrow n + d $ reaction
calculated with 
the chiral semilocal momentum-space regularized two-nucleon 
	potentials~\cite{SMS} at {N$^4$LO$^+$} augmented by the consistently regularized 
	three-nucleon force at N$^2$LO~\cite{Maris2021} for different values of the cutoff parameter ($\Lambda)$.
        We compare four different calculations: 
	(1) symmetrized plane wave with the single-nucleon and two-nucleon 
	parts in the transition operator and three-nucleon force effects included 
	in the initial three-nucleon bound state (PWIAS--(SN+2N)--(2NF+3NF)),
	(2) calculation with the initial and final states calculated with the same 
	Hamiltonian comprising two- and three-nucleon forces but retaining 
	only the single-nucleon contribution in the transition operator
	(Full--SN--(2NF+3NF)),
	(3) calculation with the initial and final states calculated with the same 
	Hamiltonian comprising only two-nucleon forces and including the single-nucleon
	and two-nucleon parts in the transition operator (Full--(SN+2N)--2NF), and
	(4) calculation with the initial and final states calculated with the
	two- and three-nucleon forces and the complete transition operator (Full--(SN+2N)--(2NF+3NF)).}
\begin{tabular}{|c|c|c|c|c|}
\hline\hline
	& \multicolumn{4}{l|}{Absorption rate $\Gamma_{nd}$ in 10$^{15}$ s$^{-1}$} \\ 
                    \cline{2-5}
\hline
	$\Lambda$ (MeV) & calc. (1) & calc. (2) & calc. (3) & calc. (4) \\ 
\hline
  400   &  8.3158     & 0.0172  &  3.6566   & 3.028  \\ 
  450   &  6.6961     & 0.0231  &  2.5466   & 2.089 \\ 
  500   &  5.4398     & 0.0666  &  1.9909   & 1.595 \\ 
  550   &  4.6015     & 0.1840  &  1.8029   & 1.371 \\
\hline\hline
\end{tabular}
\label{tablend}
\end{table}

\section{Results for the 
$\pi^- + {^3{\rm He}} \rightarrow p + n + n $ 
and
$\pi^- + {^3{\rm H}} \rightarrow n + n + n $ 
reactions}
\label{section4}

\subsection{$\pi^- + {^3{\rm He}} \rightarrow p + n + n $}

The kinematics of the three-body reactions in the center-of-mass frame 
requires in the unpolarized case two independent variables and some choices 
are of special importance. We start with the one employed recently in muon capture 
and in pion radiative capture reactions with trinucleons~\cite{PRC90.024001,PRC94.034002,PRC98.054001}, which utilizes the fact that the nuclear matrix element $N_{pnn}$ is given in terms of the Jacobi momenta. 
For the $ \pi^- + {^3{\rm He}} \rightarrow p + n + n $ reaction
this key quantity is 
\begin{eqnarray}  
N_{pnn} (m_1, m_2, m_3 , m_{^3{\rm He}}  \, )  \, \equiv \, 
{{}^{(-)}\BA{\Psi_{pnn}  \, 
	m_1 \, m_2 \, m_3 \,
	{\bf P}_f=0 
	}}\, 
	\rho_{3N}
\, \KT{\Psi_{^3{\rm He}} \, m_{^3{\rm He}} \, {\bf P}_i=0 \, 
	} 
\label{npnn}
\end{eqnarray}
and we use the following formula for the total absorption rate $\Gamma_{pnn}$:
\begin{eqnarray}
	\Gamma_{pnn} = 
   {\cal R} \, \frac{ 16 \, \left( \alpha\, M^\prime_{^3{\rm He}} \, \right)^3 \, c\, M}{ 9 M_{\pi^-} }
	  \int d {\bf\hat q} \,
	  \int\limits_{0}^{2 \pi} d \phi_{p} \,
          \int\limits_{0}^{\pi} d \theta_{p} \sin \theta_{p} \, \nonumber \\
          \times 
          \int\limits_0^{p_{max}} \, dp p^2  \,
	  \sqrt{\frac43 \left( M E_{pq} - p^2  \right)} \,
	  \frac12 \, 
	 \sum\limits_{m_1, m_2, m_3, m_{^3{\rm He}}} 
	 \left| 
	 N_{pnn} (m_1, m_2, m_3, m_{^3{\rm He}} \, ) \, 
	 \right|^2  \, ,   
\label{gpnn}
\end{eqnarray}  
where the internal energy of the final 3N state $E_{pq}$ is expressed in terms of the Jacobi relative momenta ${\bf p}$ and ${\bf q}$ 
\begin{eqnarray}
{\bf p}  \equiv \frac12 \left(  {\bf p}_2 - {\bf p}_3 \,  \right) \, , \nonumber \\
{\bf q}  \equiv \frac23 \left(  {\bf p}_1 - \frac12 \left(  {\bf p}_2 +  {\bf p}_3 \,  \right) \,  \right) \,
	= {\bf p}_1 \, ,
\label{pq}
\end{eqnarray} 
taking the following form
\begin{eqnarray}
M_\pi + M_{^3{\rm He}} 
\approx 3 M + \frac{ {\bf p}^{\, 2}} {M} + \frac34 \frac{ {\bf q}^{\, 2}} {M} 
\equiv 3 M + E_{pq}  \, 
= 3 M + \frac{ p_{max}^{\, 2}} {M} \,
= 3 M + \frac34 \frac{ q_{max}^{\, 2}} {M} \, .
\label{epq}
\end{eqnarray} 
For the unpolarized case we choose ${\bf\hat q} = {\bf\hat z}$ and $\phi_{p}=0$,
so the triple integral $ \int d {\bf\hat q} \, \int\limits_{0}^{2 \pi} d \phi_{p}$
yields a factor of $ 8 \pi^2$. 

Our predictions for $\Gamma_{pnn}$ are given in Table~\ref{tablepnn} for the same four 
types of calculations as in Table~\ref{tablend}.
The crucial role played by the two-nucleon part of the transition operator is proven also for this process (calculations (2) vs. (4)). Final state interactions lower the rates by a factor of 2.5-2.8 (calculations (1) vs. (4)). The three-nucleon force effects are now much smaller and reach 5.5~\% for $\Lambda$= 550 MeV (calculations (3) vs. (4)). The relative spread of the results obtained with different $\Lambda$ values is smaller than for the
$\pi^- + {^3{\rm He}} \rightarrow n + d $ reaction and is further reduced by more than two if the prediction for the smallest $\Lambda$= 400 MeV is dropped. 

\begin{table}
\caption{Absorption rates 
for the $\pi^- + {^3{\rm He}} \rightarrow p + n + n $ reaction
calculated with the same forces
	and with the same four types of dynamics as in the case of $\Gamma_{nd}$ in Table~\ref{tablend}.}
\begin{tabular}{|c|c|c|c|c|}
\hline\hline
	& \multicolumn{4}{l|}{Absorption rate $\Gamma_{pnn}$ in 10$^{15}$ s$^{-1}$} \\ 
                    \cline{2-5}
\hline
	$\Lambda$ (MeV) & calc. (1) & calc. (2) & calc. (3) & calc. (4) \\ 
\hline
 400   & 38.378     & 0.675  &  16.346   & 15.686 \\ 
 450   & 35.212     & 0.612  &  13.237   & 12.733 \\ 
 500   & 32.343     & 0.601  &  11.849   & 11.367 \\ 
 550   & 30.170     & 0.650  &  12.039   & 11.421 \\
\hline\hline
\end{tabular}
\label{tablepnn}
\end{table}

Using the relation between the magnitudes of the Jacobi momenta
from Eq.~(\ref{epq}) we arrive at a new formula for $\Gamma_{pnn}$ 
\begin{eqnarray}
        \Gamma_{pnn} = 
    {\cal R} \, 
    \frac{32 \, \pi^2 \, \left( \alpha\, M^\prime_{^3{\rm He}} \, \right)^3 \, c \, M\,  }{ 3 M_{\pi^-} }
          \int\limits_{0}^{\pi} d \theta_{p} \sin \theta_{p} \, \nonumber \\
          \times 
          \int\limits_0^{q_{max}} \, dq q^2  \,
	  \sqrt{M E_{pq} - \frac34 q^2 } \, \,
                \frac12 \, 
         \sum\limits_{m_1, m_2, m_3, m_{^3{\rm He}}} 
         \left| 
         N_{pnn} (m_1, m_2, m_3, m_{^3{\rm He}} \, ) \, 
         \right|^2  \, ,   
\label{gpnn.2}
\end{eqnarray}
which is a good starting point to calculate the differential absorption rates
$ \frac{ d \Gamma_{pnn} } {d E_1}$, where nucleon~1 can be a proton or a neutron.
Namely one simply reads out 
\begin{eqnarray}
	\frac{ d \Gamma_{pnn} }{ d q}  = 
	\frac{ d \Gamma_{pnn} }{ d p_1}  = 
                 {\cal R} \, 
                  \frac{32 \, \pi^2 \, \left( \alpha\, M^\prime_{^3{\rm He}} \, \right)^3 \, c \, M\,  }{ 3 M_{\pi^-} } \, q^2 \,  \sqrt{M E_{pq} - \frac34 q^2 }  \nonumber \\
                  \times
          \int\limits_{0}^{\pi} d \theta_{p} \sin \theta_{p} \, 
                  \frac12 \, 
         \sum\limits_{m_1, m_2, m_3, m_{^3{\rm He}}} 
         \left| 
         N_{pnn} (m_1, m_2, m_3, m_{^3{\rm He}} \, ) \, 
         \right|^2  \, .   
\label{gpnn.3}
\end{eqnarray}

\begin{figure}
\includegraphics[width=0.9\linewidth]{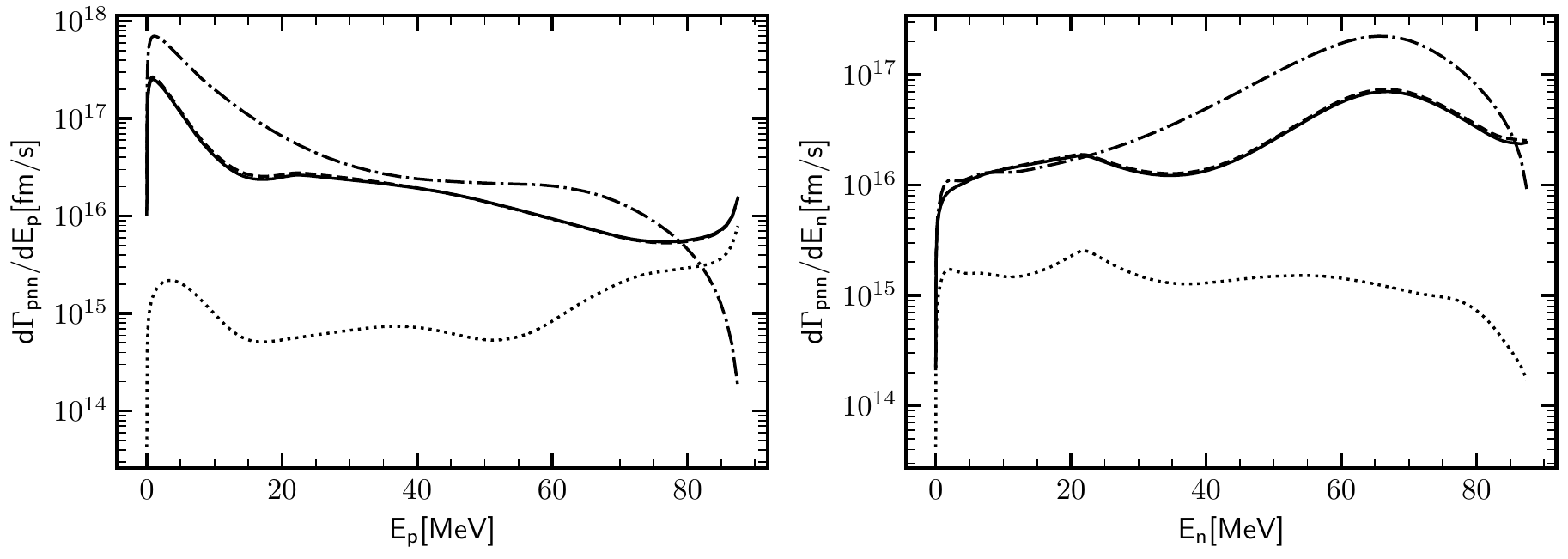}
\caption{Differential absorption rates $ {d\Gamma_{pnn} }/ {dE_p} $ (left)
and  $ {d\Gamma_{pnn} }/ {dE_n} $ (right)
	for the $\pi^- + {^3{\rm He}} \rightarrow p + n + n $ process,
calculated using the SMS chiral potential at {N$^4$LO$^+$} with $\Lambda$= 450 MeV
and with the four different computational setups defined above:
"PWIAS--(SN+2N)--(2NF+3NF)" (dash-dotted line),
"Full--SN--(2NF+3NF)" (dotted line),
"Full--(SN+2N)--2NF" (dashed line),
"Full--(SN+2N)--(2NF+3NF)" (solid line).
\label{fig3HedGdq}}
\end{figure}

In Fig.~\ref{fig3HedGdq} we show our predictions
for the single-nucleon spectra 
$\frac{ d \Gamma_{pnn} }{ d E_1}  = \frac{M}{p_1} \,\frac{ d \Gamma_{pnn} }{ d p_1}$
obtained with the four types of dynamics to show the role of the 2N contributions
in the transition operator $\rho_{3N}$ and study the 3N force effects.
Similarly as in Tables~\ref{tablend} and~\ref{tablepnn}, we see in Fig.~\ref{fig3HedGdq} that
calculations employing only the single-nucleon pion absorption operator yield much lower values of
the absorption rate. It is also visible that the rescattering part of the nuclear matrix element 
plays an important role as its inclusion reduces significantly the PWIAS predictions for $\frac{ d \Gamma_{pnn} }{ d E_p}$
as well as for  $\frac{ d \Gamma_{pnn} }{ d E_n}$ and changes their shape, making it more complicated.
In particular the PWIAS results do not exhibit any enhancement in the vicinity of the maximal proton energy.
On the other hand 3N force effects are rather small and are hardly visible on a logarithmic 
scale -- the dashed and solid lines practically overlap.

The exclusive results for the $ \pi^- + {^3{\rm He}} \rightarrow p + n + n $ reaction 
can be demonstrated in the form of double differential absorption rates. 
One of the possible choices is $ {d^2\Gamma_{pnn} }/ \left( {d E_1 \, d E_2} \right) $
calculated as 
\begin{eqnarray}
	\frac{d^2\Gamma_{pnn} }{ {d E_1 \, d E_2}  }\, = \,
                 {\cal R} \, 
                  \frac { 64\, \pi^2 \, \left( \alpha \, M^\prime_{^3{\rm He}} \, \right)^3 \, c\, M^3 } { 3 M_{\pi^-}   } \,
         \nonumber \\
         \times \,
          \frac12 \, 
         \sum\limits_{m_1, m_2, m_3, m_{^3{\rm He}}} 
         \left| 
         N_{pnn} (m_1, m_2, m_3, m_{^3{\rm He}} \, ) \, 
         \right|^2  \, ,   
\label{gpnn.4}
\end{eqnarray}
in the $(E_1, E_2)$ domain, where 
$ -1 \le \frac{E - 2 E_1 - 2 E_2 }{ 2 \sqrt{ E_1 \, E_2} } \le 1 $.
Here $E \equiv E_{pq}$ is the total kinetic 3N energy 
\begin{eqnarray}
	E = E_1 + E_2 + E_3 
	\, .
\label{E}
\end{eqnarray}
We can also introduce dimensionless variables $x$ and $y$ as
\begin{eqnarray}
	x & = & \sqrt{3} \, ( E_1 + 2 E_2 - E ) / E \, , \nonumber \\
	y & = &  ( 3 E_1 - E ) / E \, , 
\label{xy}
\end{eqnarray}
restricted to the disk $ r^2 \equiv x^2 + y^2 \le 1 $
and evaluate 
$ {d^2\Gamma_{pnn} }/ \left( {d x \, d y} \right) $
or (using polar coordinates)
$ {d^2\Gamma_{pnn} }/ \left( {d r \, d \phi} \right)$.
Such forms were used for example in Ref.~\cite{Gotta1995}.

\begin{figure}
\includegraphics[width=0.8\linewidth]{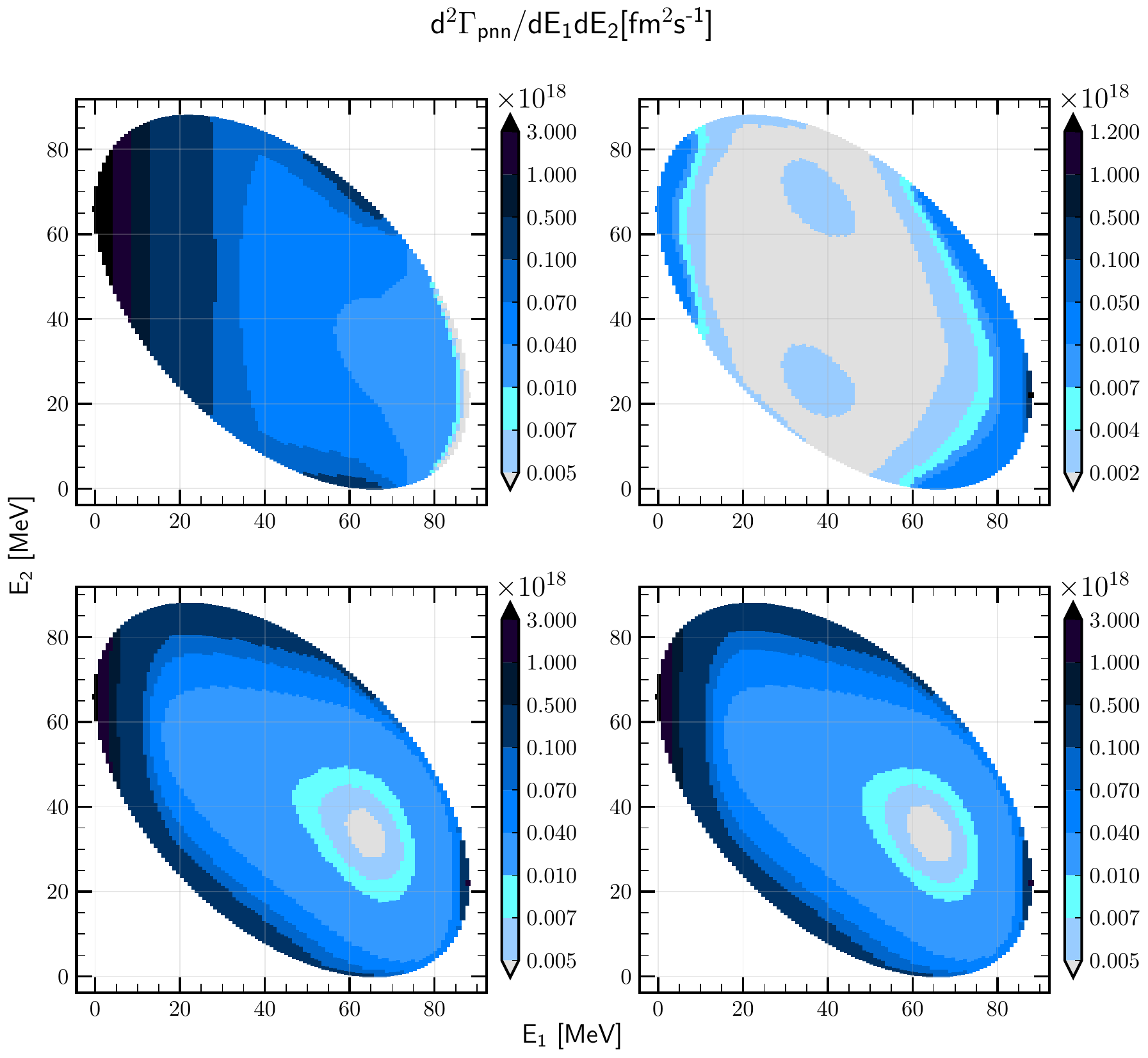}
\caption{Intensity plots for the double 
differential absorption rates 
	$ {d^2\Gamma_{pnn} }/\left( dE_1 dE_2 \right)  $ 
for the $\pi^- + {^3{\rm He}} \rightarrow p + n + n $ process,
obtained using the SMS chiral potential at {N$^4$LO$^+$} with $\Lambda$= 450 MeV
in the four computational setups defined above:
"PWIAS--(SN+2N)--(2NF+3NF)" (top left),
"Full--SN--(2NF+3NF)" (top right),
"Full--(SN+2N)--2NF" (bottom left),
"Full--(SN+2N)--(2NF+3NF)" (bottom right).
Nucleon~1 is a proton.
\label{fig3Hed2GdE1dE2}}
\end{figure}

In Fig.~\ref{fig3Hed2GdE1dE2} we show the $ {d^2\Gamma_{pnn} }/\left( dE_1 dE_2 \right)  $ obtained within the same
dynamical models as used in Fig.~\ref{fig3HedGdq} for the  cutoff value $\Lambda=450$~MeV. Our most advanced prediction "Full--(SN+2N)--(2NF+3NF)"
is given in the right bottom panel. The dominant contributions to this double differential
capture rate stem from configurations close to the borders of the kinematically allowed region.
This could be expected as some special kinematical configurations are located there. For the nucleon-induced deuteron breakup process, where there are three free nucleons
in the final state, the differential cross section is enhanced in specific kinematical configurations~\cite{Glocklephysrep}. We observe the same phenomenon in the pion absorption differential rates discussed here.
The first type of important configurations is related to a strong final state interaction between two nucleons emerging with zero or very small relative energy.
In the neutron-neutron final state interaction (FSI(nn)) configuration the proton (particle 1) gets two-thirds of the available kinetic energy and each neutron one-sixth of it, with $ {\bf p}_1 = -2{\bf p}_2 = -2{\bf p}_3$. Rescattering effects at the FSI(nn) kinematics increase the capture rates significantly, from 9.7$\times$10$^{14}$~fm$^2$~s$^{-1}$ to 2.36$\times$10$^{18}$~fm$^2$~s$^{-1}$.
This is seen when comparing left upper and right bottom plots, for which predictions differ only in taking FSI into account. 
A rapid drop of the $ {d\Gamma_{pnn} }/ dE_p  $ at high values of $E_1$ also appears on Fig.~\ref{fig3HedGdq} (left). 
The change of the capture rates for the two proton-neutron final state interaction kinematical configurations (FSI(pn)) located on the graphs at maximal E$_2$ and  the diagonal E$_1$=E$_2$ for the lowest allowed energies ($\approx 23$~MeV) is smaller. For example, for FSI(pn) at E$_2$=87.46~MeV "PWIAS--(SN+2N)--(2NF+3NF)" capture rate $ {d^2\Gamma_{pnn} }/\left( dE_1 dE_2 \right)  $ is 1.5$\times$10$^{17}$~fm$^2$~s$^{-1}$ while the "Full--(SN+2N)--(2NF+3NF)" capture rate amounts to 4.48$\times$10$^{17}$~fm$^2$~s$^{-1}$.
The other interesting kinematical configurations located on the border of kinematically allowed region correspond to quasi-free scattering (QFS), where two nucleons share equally the absorbed energy, while the third (spectator) nucleon remains at rest. Obviously, two such configurations are located at E$_1$=0~MeV (QFS(nn)) and at E$_2$=0 (QFS(pn)). The remaining QFS(pn) configuration is
located at the diagonal E$_1$=E$_2$ for the highest allowed energies ($\approx 68$~MeV). Here effects of the FSI are less pronounced and change the absorption rates up to 17\%. The absorption rate $ {d^2\Gamma_{pnn} }/\left( dE_1 dE_2 \right)  $ reaches its maximum just for the QFS(nn) configuration.

A comparison of the results shown in the right column of Fig.~\ref{fig3Hed2GdE1dE2} brings information on
the role of the two-nucleon components of the pion absorption operator. The capture rates obtained with the single nucleon absorption operator only are definitely smaller than those obtained in the "Full--(SN+2N)--(2NF+3NF)" model, and the "Full--SN--(2NF+3NF)" total capture rate 
receives nearly all non-negligible contributions from configurations with low or high E$_1$ energies. 

Finally, by comparing two plots in the bottom row of Fig.~\ref{fig3Hed2GdE1dE2} we demonstrate that three-body forces do not play a significant role in this process. Since inclusion of a three-nucleon force in the bound state calculations changes the $^3$He wave function, this comparison might suggest that the $\pi^- + {^3{\rm He}} \rightarrow p + n + n $ reaction is rather insensitive to details of the nuclear bound state.

\begin{figure}
\includegraphics[width=0.83\linewidth]{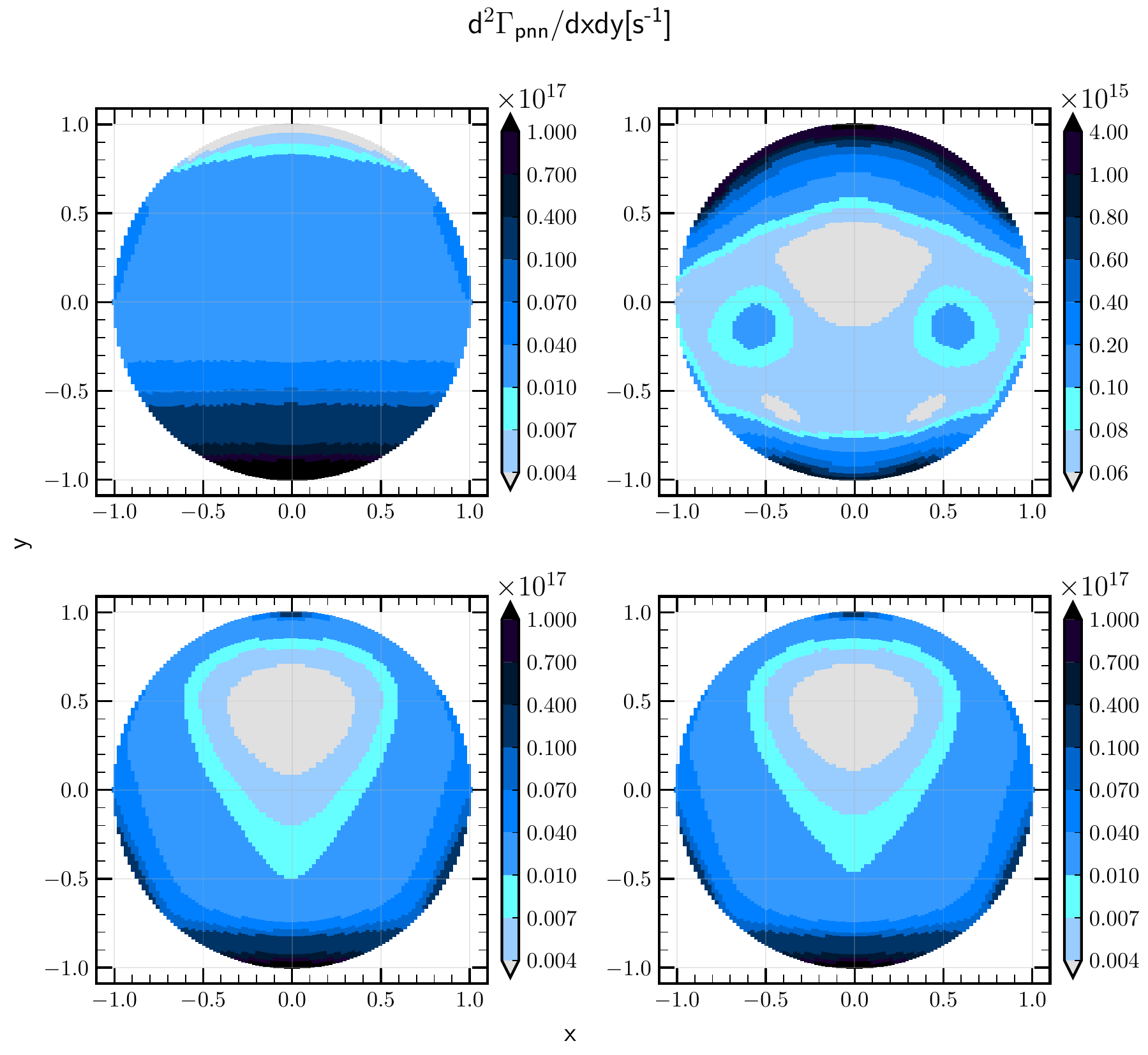}
\caption{Intensity plots for the double 
differential absorption rates 
	$ {d^2\Gamma_{pnn} }/\left( dx dy \right)  $ 
for the $\pi^- + {^3{\rm He}} \rightarrow p + n + n $ process,
obtained using the SMS potential at {N$^4$LO$^+$} with $\Lambda$= 450 MeV
in the four different computations defined above:
PWIAS--(SN+2N)--(2NF+3NF)" (top left),
"Full--SN--(2NF+3NF)" (top right),
"Full--(SN+2N)--2NF" (bottom left),
"Full--(SN+2N)--(2NF+3NF)" (bottom right).
Nucleon~1 is a proton.
\label{fig3Hed2Gdxdy}}
\end{figure}

Obviously, the same observations arise from Fig.~\ref{fig3Hed2Gdxdy}, where we show the double differential absorption rates $ {d^2\Gamma_{pnn} }/\left( dx dy \right)$. Here, the FSI(nn) configuration is placed at 
$(x,y)=(0,1)$ and the two FSI(pn) configurations are positioned on the border of the kinematically allowed region at an angle of 120$^{\circ}$ from FSI(nn) point symmetrically: clockwise and counterclockwise. In the $(x,y)$ representation the QFS(nn) configuration is located at $(x,y)=(0,-1)$ and again the angles corresponding to the two QFS(pn) configurations differ from the QFS(nn) angle by $\pm$120$^{\circ}$. 
The results in the right bottom panel of Fig.~\ref{fig3Hed2Gdxdy} highlight even better  the importance of the above mentioned special configurations than the representation used in Fig.~\ref{fig3Hed2GdE1dE2}. 
The "Full--SN--(2NF+3NF)" model delivers the absorption rate approximately two orders of magnitude smaller than the predictions based on the single-nucleon absorption operator supplemented by two-nucleon operators.
The expected left-right symmetry seen in all panels of Fig.~\ref{fig3Hed2Gdxdy} confirms the high accuracy of our numerical methods. 

\begin{figure}
\includegraphics[width=0.7\linewidth]{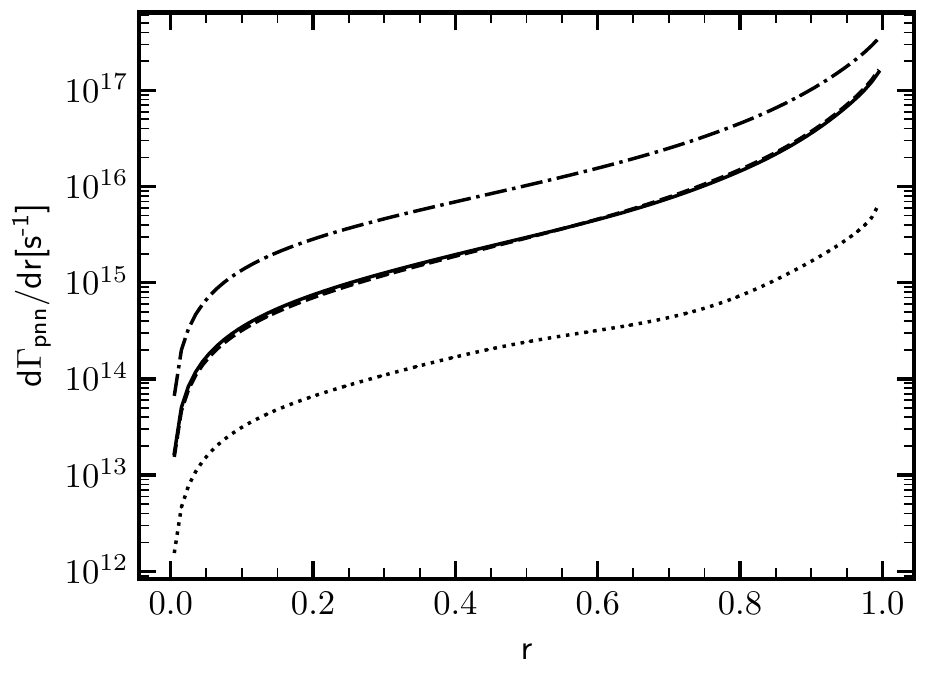}
\caption{Differential absorption rate $ {d\Gamma_{pnn} }/dr  $
for the $\pi^- + {^3{\rm He}} \rightarrow p + n + n $ process,
calculated using the SMS potential at {N$^4$LO$^+$} with $\Lambda$= 450 MeV
and with the four dynamical setups defined above:
"PWIAS--(SN+2N)--(2NF+3NF)" (dash-dotted line),
"Full--SN--(2NF+3NF)" (dotted line),
"Full--(SN+2N)--2NF" (dashed line),
"Full--(SN+2N)--(2NF+3NF)" (solid line).
Nucleon~1 is a proton.
\label{fig3HedGdr}}
\end{figure}

Since the intensity plots (Figs.~\ref{fig3Hed2GdE1dE2} and \ref{fig3Hed2Gdxdy}) 
suggest that $\Gamma_{pnn}$ receives dominant contributions from the regions in the vicinity 
of the boundaries, in Fig.~\ref{fig3HedGdr} we show also the differential 
absorption rate $ {d\Gamma_{pnn} }/ dr $. Indeed, for all approaches, including even the PWIAS ones
the dominant contribution to the total absorption rate $\Gamma_{pnn}$ arises from the narrow ring with 
$r > 0.9$ while the contributions coming from the central region ($r \lesssim 0.1$) are 
four orders of magnitude smaller. Note however that, in accordance with Fig.~\ref{fig3Hed2Gdxdy}, for the "PWIAS--(SN+2N)--(2NF+3NF)", the "Full--SN--(2NF+3NF)",
and the "Full--(SN+2N)--(2NF+3NF)" approaches the highest absorption rates are located at different parts of the ring $ r> 0.9 $.

\begin{figure}
\includegraphics[width=0.7\linewidth]{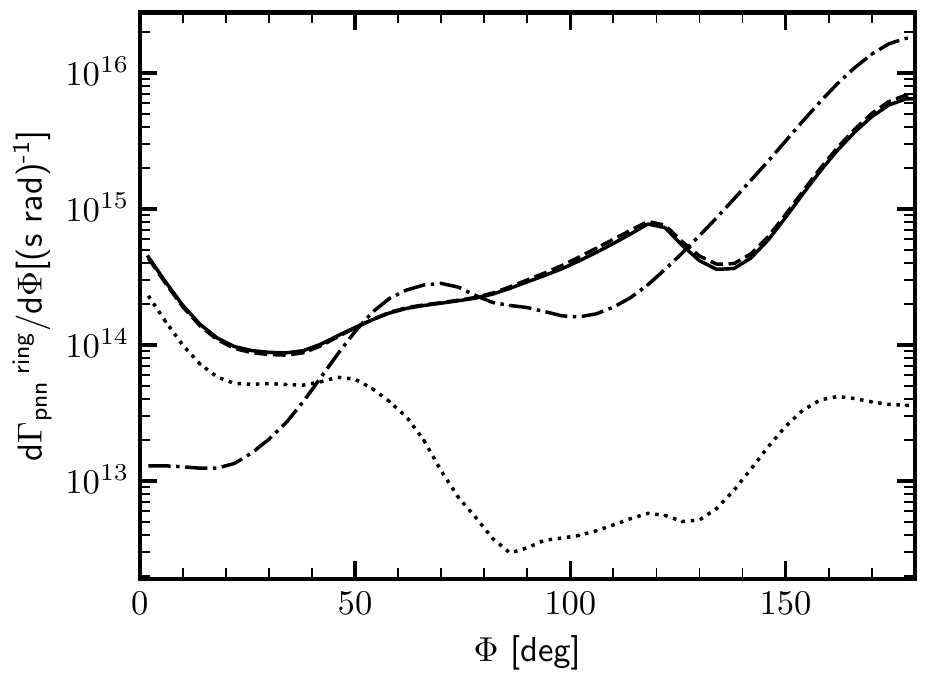}
\caption{Angular distribution of the differential absorption rate 
	$ {d\Gamma_{pnn}^{\rm ring} }/d \Phi$
for the $\pi^- + {^3{\rm He}} \rightarrow p + n + n $ process,
calculated using the SMS chiral potential with $\Lambda$~=~450 MeV
and with the four dynamical setups defined above:
"PWIAS--(SN+2N)--(2NF+3NF)" (dash-dotted line),
"Full--SN--(2NF+3NF)" (dotted line),
"Full--(SN+2N)--2NF" (dashed line),
"Full--(SN+2N)--(2NF+3NF)" (solid line).
Nucleon~1 is a proton and due to the symmetry of the problem
	the angular distribution is shown in the interval 
	$0 \le \Phi \le 180^\circ$.
\label{fig3HedGdphi}}
\end{figure}

In Ref.~\cite{Gotta1995} the authors considered the 
two-dimensional distribution of experimental events 
in the narrow ring near the boundary of the kinematically 
allowed region defined by $ 0.95 \le r \le 1$. 
We thus calculate the corresponding quantity 
	$ {d\Gamma_{pnn}^{\rm ring} }/d \Phi =
	\int\limits_{0.95}^1 dr \, {d\Gamma_{pnn} }/( dr d \Phi )  $
and show the results in Fig.~\ref{fig3HedGdphi}. The angle $\Phi$ is measured
clockwise from the point $(x=0, y=1)$, where $E_1$ has the maximal value 
$E_1 = \frac23 E$ and $E_2 = E_3 = \frac16 E$. 
This corresponds to the final state interaction kinematics where ${\bf p}_2 = {\bf p}_3 = - \frac12 {\bf p}_1$.
The angular dependence of the $ {d\Gamma_{pnn}^{\rm ring} }/d \Phi$ changes with 
the dynamical approach used. Specifically, the model employing only the single nucleon pion absorption operator
predicts the dominant contribution from the vicinity of the FSI(nn) configuration (low $\Phi$), while for the PWIAS model the absorption rate in question has minimal values at low $\Phi$ 
and rises more than 1000 times to achieve a maximum at $\Phi=180^{\circ}$ (region close to
the QFS(pn) configuration). For the two remaining models the  $ {d\Gamma_{pnn}^{\rm ring} }/d \Phi$ 
has maxima both at low and high $\Phi$, however the latter maximum is around one order of magnitude higher.
A local maximum seen at $\Phi=120^{\circ}$ is related to the FSI(pn) region.

In Ref.~\cite{Gotta1995} the authors consider four regions, 
$I_1$, 
$I_2$, 
$I_3$, and
$I_4$, 
in the phase space of the $\pi^- + {^3{\rm He}} \rightarrow p + n + n $ process,
which are defined in Fig.~14 and Table~III of their paper.
The regions were determined in terms of the $\Phi$ angles of the Dalitz plot 
and we could calculate, with various dynamics, the corresponding parts 
(denoted by $\Gamma_i$) of the total absorption rates $\Gamma_{pnn}$.
Our predictions for the most consistent dynamics are presented in Table~\ref{tableregions}.

\begin{table}
\caption{Absorption rates in the four regions 
of the phase space $I_i$ defined in 
	Ref.~\cite{Gotta1995} 
for the $\pi^- + {^3{\rm He}} \rightarrow p + n + n $ reaction
calculated with 
the chiral semilocal momentum-space regularized two-nucleon 
	potentials~\cite{SMS} at {N$^4$LO$^+$} augmented by the consistently regularized 
	three-nucleon force at N$^2$LO~\cite{Maris2021} with different values of the cutoff parameter $\Lambda$. 
	The rates are determined with the two- and three-nucleon forces 
	and the complete transition operator (Full--(SN+2N)--(2NF+3NF)).}
\begin{tabular}{|c|c|c|c|c|}
\hline\hline
	& \multicolumn{4}{l|}{Absorption rates $\Gamma_{i}$ in 10$^{15}$ s$^{-1}$} \\ 
                    \cline{2-5}
\hline
	$\Lambda$ (MeV)                             &      $I_1$ &      $I_2$ &      $I_3$ &      $I_4$ \\ 
\hline
 400   &  9.463  & 1.789  & 0.340  & 0.185 \\ 
 450   &  7.614  & 1.452  & 0.308  & 0.179 \\ 
 500   &  6.758  & 1.298  & 0.299  & 0.176 \\ 
 550   &  6.826  & 1.302  & 0.309  & 0.173 \\
\hline\hline
\end{tabular}
\label{tableregions}
\end{table}

In order to compare our results with \cite{Gotta1995} and
avoid normalization issues we put in Table~\ref{tableregions_norm} the same results  as in Table~\ref{tableregions} 
but normalized by the sum over all regions $I_i$. The same normalization is applied to the experimental results from the paper and listed in the same table.
In such a way we do not compare the absolute values of the partial absorption rates, but their relative values. 
One can see that most of the regions have values comparable to the experimental results.
The biggest inconsistency is found for the region $I_2$, while the best agreement emerges for the region $I_3$. 
For the normalized values the dependence on the cutoff value $\Lambda$ gets weaker, which means that the choice of the cutoff affects more strongly the 
absolute values of the absorption rate than its distribution over the available phase space.

\begin{table}
\caption{The same as
in \ref{tableregions} but all the values are normalized by the sum $I_1 + I_2 + I_3 + I_4$.
The experimental results from	Ref.~\cite{Gotta1995} shown in the table are also normalized in the same way.}
\begin{tabular}{|c|c|c|c|c|}
\hline\hline
	& \multicolumn{4}{c|}{Normalized absorption rates $\Gamma_{i}$} \\ 
                    \cline{2-5}
\hline
	$\Lambda$ (MeV)                             &      $I_1$ &      $I_2$ &      $I_3$ &      $I_4$ \\ 
\hline
 400   &  \;\;0.804\;\; &\;\; 0.152\;\;  &\;\;\ 0.029\;\;  & \;\;0.016\;\; \\
 450   &  0.797  & 0.152  & 0.032  & 0.019 \\
 500   &  0.792  & 0.152  & 0.035  & 0.021 \\
 550   &  0.793  & 0.151  & 0.036  & 0.020  \\
\hline\hline
Gotta \textit{et al.} \cite{Gotta1995} &    0.844& 0.099& 0.033& 0.023 \\
\hline\hline
\end{tabular}
\label{tableregions_norm}
\end{table}

Our results for the $\Gamma_{nd}$ and $\Gamma_{pnn}$ rates 
were calculated by explicit summation (integration) over the two- and three-body 
nuclear states. We can also employ the method which uses closure \cite{PRC98.054001} and allows us 
to determine the total absorption rate, which should agree with $\Gamma_{nd} + \Gamma_{pnn}$.
This important numerical check for our most complete dynamics is shown in Table~\ref{tableclosure3He}.
The obtained agreement is in general very good and the differences of predictions
obtained within the two methods remain below 1.5\%. Such an accuracy level of our numerical calculations was estimated in \cite{Glocklephysrep}. 

\begin{table}
\caption{Absorption rates 
	$\Gamma_{nd}$, 
	$\Gamma_{pnn}$, 
	and their sum $\Gamma_{nd} + \Gamma_{pnn}$,
	compared to the result $\Gamma_{^3{\rm He}}^{\rm closure}$, 
	obtained using the closure method described in Ref.~\cite{PRC98.054001}
for negative pion absorption in $^3$He calculated with 
the chiral semilocal momentum-space regularized two-nucleon 
	potentials~\cite{SMS} at {N$^4$LO$^+$} augmented by the consistently regularized 
	three-nucleon force at N$^2$LO~\cite{Maris2021} with different values of the cutoff parameter $\Lambda$.
	The initial and final states are computed with the
	two- and three-nucleon forces and the complete transition operator 
	is employed (Full--(SN+2N)--(2NF+3NF)).}
\begin{tabular}{|c|c|c|c|c|}
\hline\hline
	& \multicolumn{4}{l|}{Absorption rates in 10$^{15}$ s$^{-1}$} \\ \cline{2-5}
 $\Lambda$ (MeV) & $\Gamma_{nd}$  & $\Gamma_{pnn}$  & $\Gamma_{nd} + \Gamma_{pnn}$ & $\Gamma_{^3{\rm He}}^{\rm closure}$ \\ 
\hline
 400   &  3.028    &  15.686 & 18.714    & 18.490  \\ 
 450   &  2.089    &  12.733 & 14.822    & 14.621 \\ 
 500   &  1.595    &  11.367 & 12.961    & 12.772 \\ 
 550   &  1.371    &  11.421 & 12.792    & 12.598 \\
\hline\hline
\end{tabular}
\label{tableclosure3He}
\end{table}

The results shown in  Table~\ref{tableclosure3He} can be summarized as follows: 
\begin{eqnarray}\label{Gam_nd_our}
\Gamma_{nd} & =  & (2.0^{+1.0}_{-0.6}\pm 1.6)\times 10^{15} \, {\rm s^{-1}}\\ \label{Gam_pnn_our}
\Gamma_{pnn}  & =& (12.8^{+ 2.9}_{-1.4} \pm 10.2)\times 10^{15} \, {\rm s^{-1}}\\ \label{Gam_nd_pnn_our}
\Gamma_{nd} + \Gamma_{pnn}   & =& (14.8^{+ 3.9}_{- 2.0} \pm 11.8 )\times  10^{15} \, {\rm s^{-1}},
\end{eqnarray}
 where  the central value  and the first  (asymmetric) uncertainty correspond to the average and the spread of the results in Table~\ref{tableclosure3He} 
from the cutoff variation, while the second error is related  to the truncation of the chiral expansion at the LO-MCS. 

  Absolute rates for different absorption channels in $^3$He are also available from experiments. The broadening of the ground state level 
of the ${\pi^-}^3$He atom  measured  in the X-ray transitions $np\to 1s$  reads~\cite{Schwanner:1984sg} (see also Ref.~\cite{Mason:1980vg} for 
an older measurement)
\begin{equation}
\Gamma_{1s} = (28\pm 7) {\rm eV} = (43\pm 11) \times 10^{15} \, {\rm s^{-1}}.
\end{equation}
Further,  branching ratios   for the channels $\pi^- + {^3{\rm He}} \rightarrow n + d $ and $\pi^- + {^3{\rm He}} \rightarrow p+n+n $
were measured  to be (16 $\pm$ 2)\% \cite{Branchings3He,McCarthy:1975zzb}  and (58 $\pm$ 5)\% \cite{Branchings3He}, respectively. 
The sum of these two channels was also measured indirectly to be (68.2 $\pm$ 2.6)\%  \cite{Truoel:1974is}. Using this information, one can extract the individual and combined 
contributions  to  the absorption rate on $^3$He, which read
\begin{eqnarray}\label{Gam_He_exp}
\Gamma_{nd}^{\rm exp.} & =  & (6.8\pm 1.9)\times 10^{15} \, {\rm s^{-1}}\\
\Gamma_{pnn} ^{\rm exp.} & =& (24.7\pm 6.5)\times 10^{15} \, {\rm s^{-1}}\\
\Gamma_{nd+pnn} ^{\rm exp.} & =& (29.0 \pm 7.3)\times 10^{15} \, {\rm s^{-1}}.
\end{eqnarray}
 While our central values are clearly  smaller than the experimental one,  our results are still generally    consistent with the data within  errors.
 Future studies should show if  agreement with the data  improves when higher-order production operators are included.

\subsection{$\pi^- + {^3{\rm H}} \rightarrow n + n + n $ }


The formula for the total absorption rate $\Gamma_{nnn}$ in the case of
the $\pi^- + {^3{\rm H}} \rightarrow n + n + n $ reaction,
\begin{eqnarray}
        \Gamma_{nnn} = 
 \frac { 2\, \left( \alpha \, M^\prime_{^3{\rm H}} \, \right)^3 \, c \, M}
 { 27 M_{\pi^-}  }
          \int d {\bf\hat q} \,
          \int\limits_{0}^{2 \pi} d \phi_{p} \, 
          \int\limits_{0}^{\pi} d \theta_{p} \sin \theta_{p} \, 
          \nonumber \\
          \times 
          \int\limits_0^{p_{max}} \, dp p^2  \,
          \sqrt{\frac43 \left( M E_{pq} - p^2  \right)} \,
         \frac12 \, 
         \sum\limits_{m_1, m_2, m_3, m_{^3{\rm H}}} 
         \left| 
         N_{nnn} (m_1, m_2, m_3, m_{^3{\rm He}} \, ) \, 
         \right|^2  \, ,   
\label{gnnn}
\end{eqnarray}
is very similar to the one in Eq.~(\ref{gpnn}); obviously it contains 
the appropriate matrix element
\begin{eqnarray}  
N_{nnn} (m_1, m_2, m_3 , m_{^3{\rm H}}  \, )  \, \equiv \, 
{{}^{(-)}\BA{\Psi_{nnn}  \, 
        m_1 \, m_2 \, m_3 \,
        {\bf P}_f=0 
        }}\, 
        \rho_{3N}
\, \KT{\Psi_{^3{\rm H}} \, m_{^3{\rm H}} \, {\bf P}_i=0 
        } \, ,
\label{nnnn}
\end{eqnarray}
an altered factor $\frac12 \rightarrow \frac16$ due the three identical 
particles in the final state as well as the proper reduced mass
of the $\pi^- - {^3{\rm H}}$ system,
$ M^\prime_{^3{\rm H}}  = \frac { M_{^3{\rm H}} M_{\pi^-} } { M_{^3{\rm H}} + M_{\pi^-} }$. 
The ${\cal R}$ factor is very close to $1$ and thus unnecessary for this nucleus.

Results of our calculations for $\Gamma_{nnn}$ at {N$^4$LO$^+$} with different cutoff values and for the same four types of calculations as for $\Gamma_{pnn}$ are given in Table~\ref{tablennn}.
For all the four computations one can observe variations depending 
on the cutoff value - the maximal difference is for the "Full--SN--(2NF+3NF)" (calc. 2)
and is up to 34\%  while the smallest cutoff dependence is for 
"PWIAS--(SN+2N)--(2NF+3NF)" (calc. 1) - less then 10\%. 
For each cutoff value the changes between different types of calculations remain
similar: the rescattering part of the nuclear matrix element and the 2N contribution
to the absorption operator are clearly important
and have a very strong influence on the final value. On the other hand, inclusion of 3NF
does not change the predictions substantially. 
The summary of the complete results (calc.4) can be read as 
\begin{eqnarray}\label{Gam_nnn_our}
\Gamma_{nnn} & =  & (1.1^{ + 0.2}_{- 0.1} \pm0.9 ) \, 10^{15} {\rm s^{-1}},
\end{eqnarray}
where as before, we averaged over the four cutoff values and added the truncation uncertainty at the LO-MCS.

\begin{table}
\caption{Absorption rates 
for the $\pi^- + {^3{\rm H}} \rightarrow n + n + n $ reaction
calculated with the same combinations of 2N and 3N potentials 
	and with the same four types of dynamics as in the case of $\Gamma_{nd}$.}
\begin{tabular}{|c|c|c|c|c|}
\hline\hline
	& \multicolumn{4}{l|}{Absorption rate $\Gamma_{nnn}$ in 10$^{15}$ s$^{-1}$} \\ 
                    \cline{2-5}
\hline
	$\Lambda$ (MeV) & calc. (1) & calc. (2) & calc. (3) & calc. (4) \\ 
\hline
400 &  2.352 & 0.086 & 1.360 & 1.375 \\ 
450 &  2.264 & 0.074 & 1.103 & 1.110 \\ 
500 &  2.179 & 0.065 & 0.999 & 1.002 \\ 
550 &  2.120  & 0.057 & 1.056 & 1.061 \\
%
\hline\hline
\end{tabular}
\label{tablennn}
\end{table}

\begin{figure}
\includegraphics[width=0.7\linewidth]{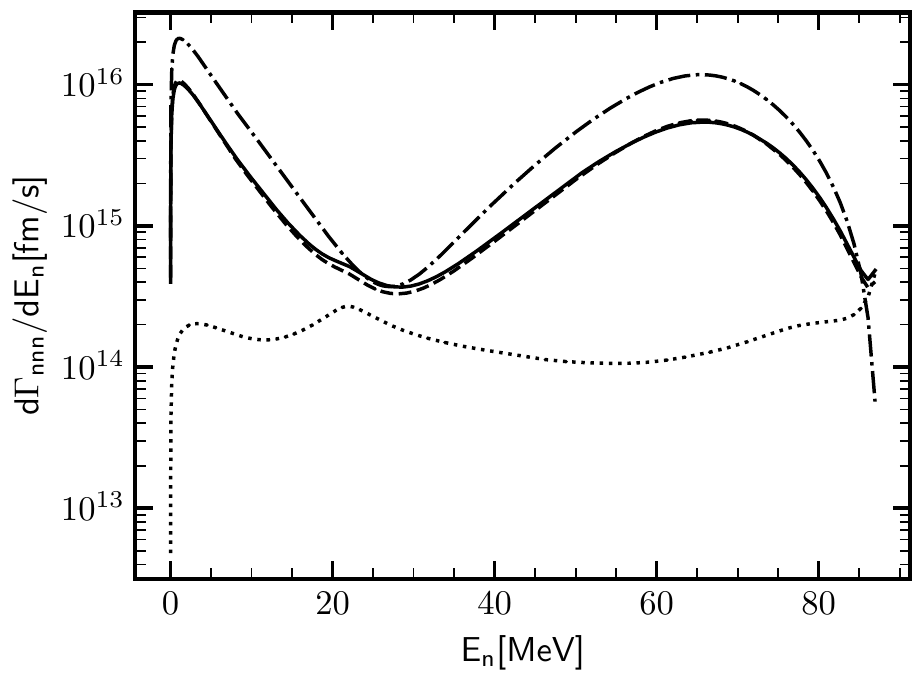}
	\caption{The same as in Fig.~\ref{fig3HedGdq} for
	the neutron spectrum 
$ {d\Gamma_{nnn} }/ {dE_n} $
        in the $\pi^- + {^3{\rm H}} \rightarrow n + n + n $ process.
\label{fig3HdGdq}}
\end{figure}

In Fig.~\ref{fig3HdGdq} we show, in analogy to Fig.~\ref{fig3HedGdq}, the differential absorption 
rates $ {d\Gamma_{nnn} }/ {dE_n} $. The structure of the pion absorption operator has a decisive influence on the observed spectrum. For most of energies inclusion of two-body absorption operators increases 
the absorption rate by about two orders of magnitude. 
The other predictions are more compatible with each other. The "PWIAS--(SN+2N)--(2NF+3NF)" results are much bigger than the others (by a factor of 2-5) and reach their minima both at the neutron's lowest and highest energy. The 
"Full--(SN+2N)--2NF" and "Full--(SN+2N)--(2NF+3NF)" predictions practically overlap. All predictions comprising the final state interactions  exhibit a sudden enhancement of the absorption rate at the right limit of the spectrum.

\begin{figure}
\includegraphics[width=0.83\linewidth]{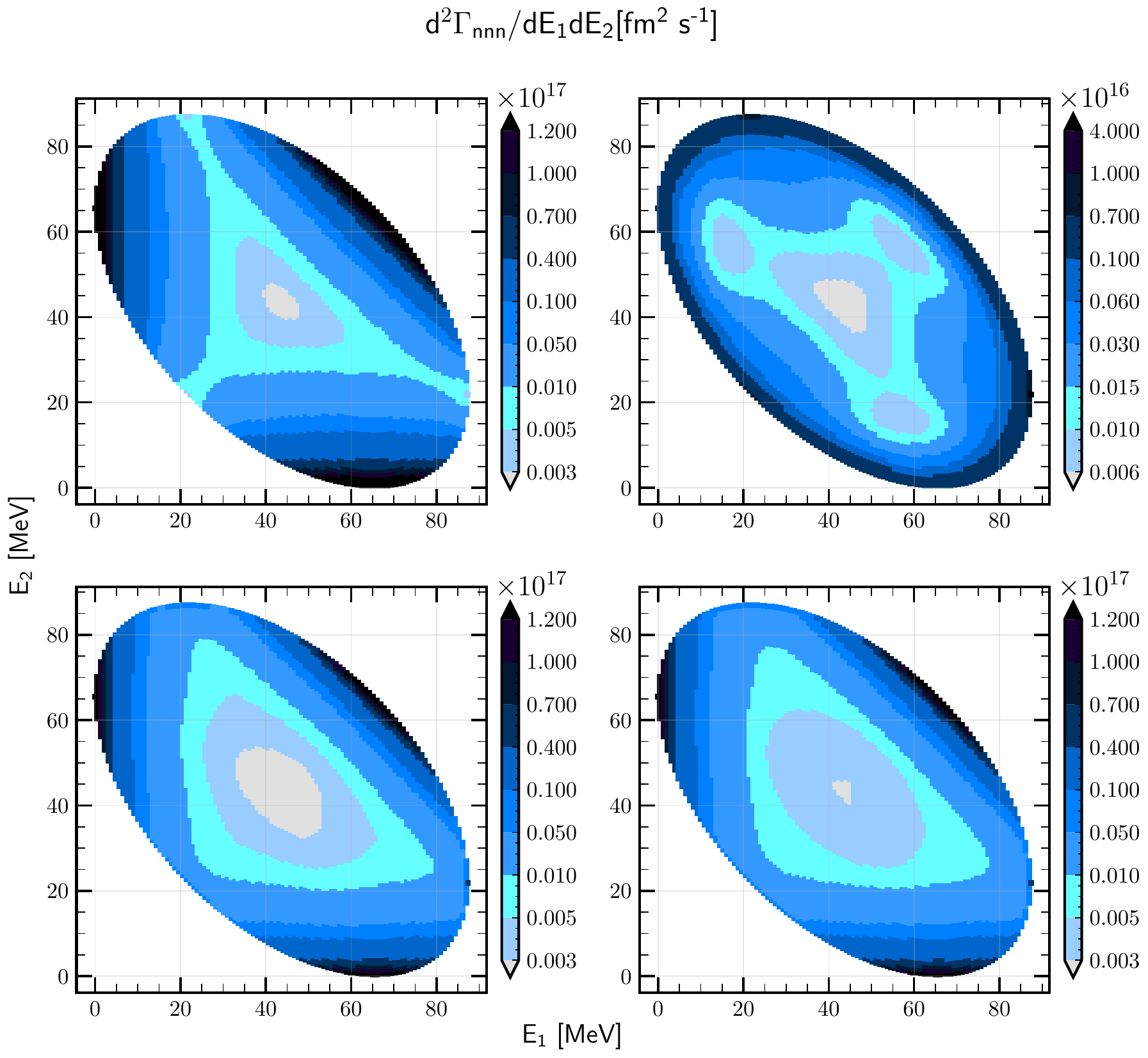}
\caption{The same as in Fig.~\ref{fig3Hed2GdE1dE2}
	for the double differential absorption rates
        $ {d^2\Gamma_{nnn} }/\left( dE_1 dE_2 \right)  $
for the $\pi^- + {^3{\rm H}} \rightarrow n + n + n $ process.
\label{fig3Hd2GdE1dE2}}
\end{figure}

\begin{figure}
\includegraphics[width=0.83\linewidth]{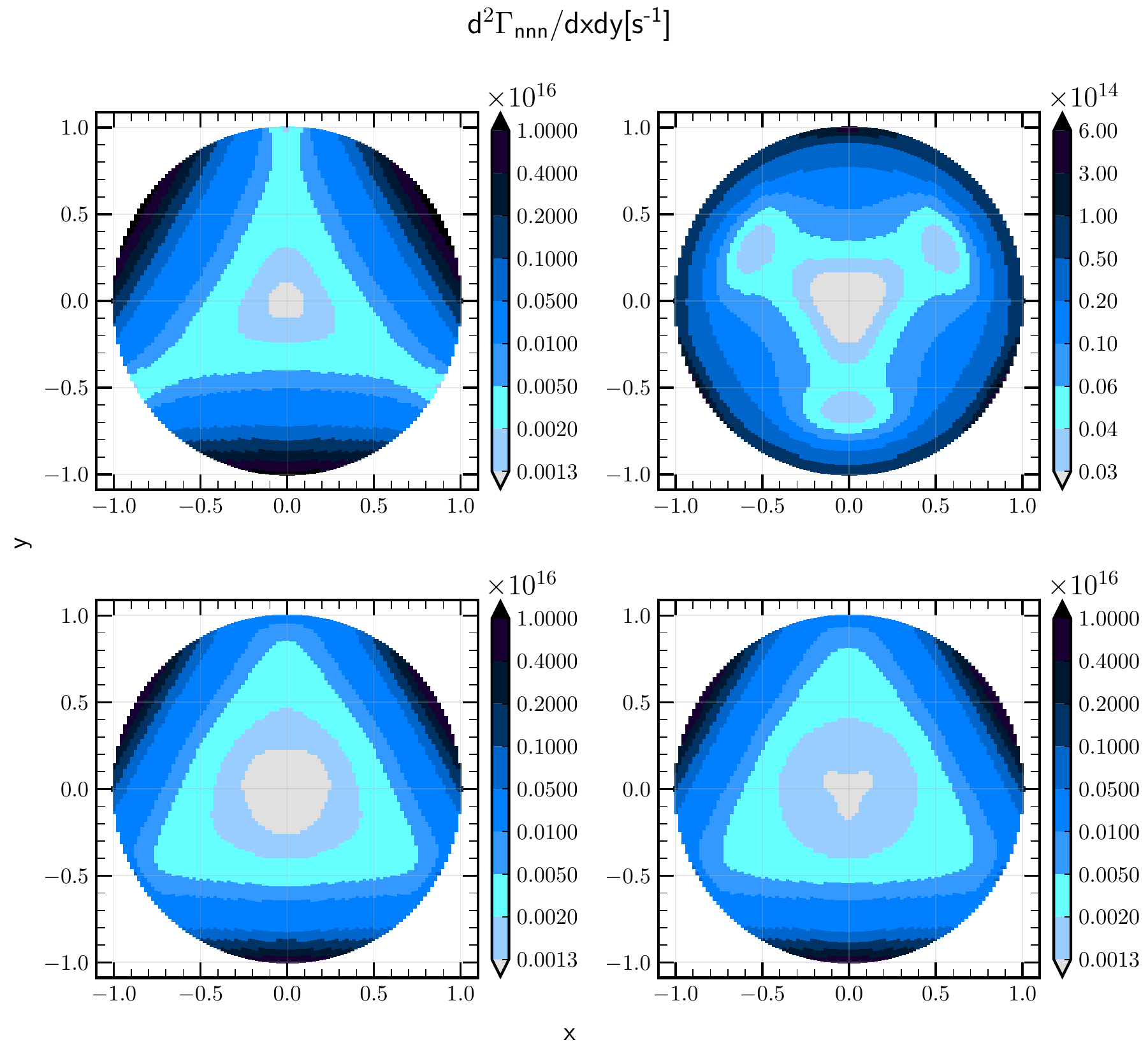}
\caption{The same as in Fig.~\ref{fig3Hed2Gdxdy}
	for the double
differential absorption rates
        $ {d^2\Gamma_{nnn} }/\left( dx dy \right)  $
for the $\pi^- + {^3{\rm H}} \rightarrow n + n + n $ process.
\label{fig3Hd2Gdxdy}}
\end{figure}

Following the path used for the pion absorption on $^3$He, in Figs.~\ref{fig3Hd2GdE1dE2} and~\ref{fig3Hd2Gdxdy}
we show the absorption rates $ {d^2\Gamma_{nnn} }/\left( dE_1 dE_2 \right)  $ and $ {d^2\Gamma_{nnn} }/\left( dx dy \right)  $, respectively. The symmetry of presented plots is, of course, due to three indistinguishable neutrons in the final state. With the exception of the "Full--SN--(2NF+3NF)" predictions, the dominant contributions
to the absorption rates come exclusively from the three QFS(nn) configurations. The inclusion of the rescattering effects or the three-nucleon interaction does not change this picture and only slightly modifies magnitudes
of the absorption rates. For the QFS(nn) the differential absorption rate $ {d^2\Gamma_{nnn} }/\left( dE_1 dE_2 \right)  $ changes from 4.3$\times$10$^{17}$ for the "PWIAS--(SN+2N)--(2NF+3NF)", 6.05$\times$10$^{16}$  for the "Full--SN--(2NF+3NF)", and 2.24$\times$10$^{17}$
for "Full--(SN+2N)--2NF" to 2.09$\times$10$^{17}$ for the most complete "Full--(SN+2N)--(2NF+3NF)" prediction (all values in fm$^2$~s$^{-1}$). In the case of the FSI(nn) configuration we observe only a moderate increase of the absorption rates when the the final state interactions are included.

\begin{figure}
\includegraphics[width=0.7\linewidth]{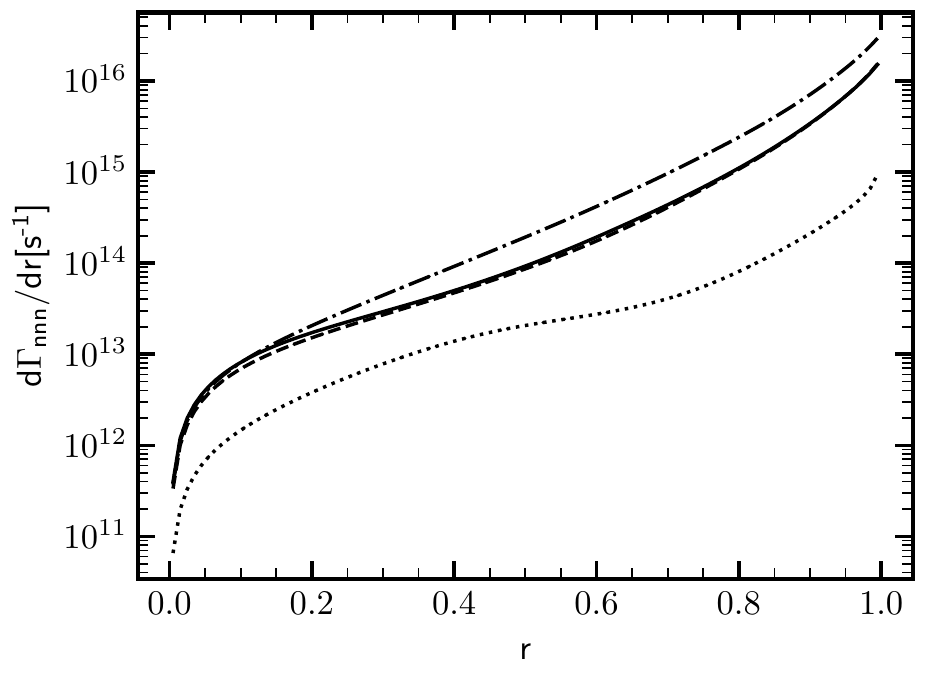}
\caption{The same as in Fig.~\ref{fig3HedGdr} 
for the $\pi^- + {^3{\rm H}} \rightarrow n + n + n $ process.
\label{fig3HdGdr}}
\end{figure}

\begin{figure}
\includegraphics[width=0.7\linewidth]{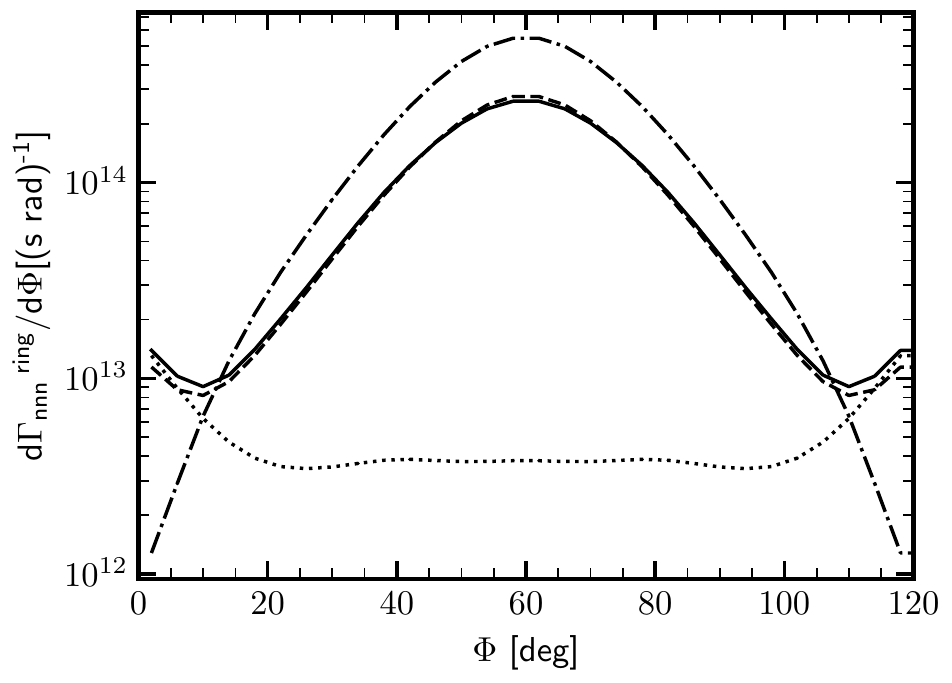}
\caption{The same as in Fig.~\ref{fig3HedGdphi} 
for the $\pi^- + {^3{\rm H}} \rightarrow n + n + n $ process,
The angular distribution inside the ring is shown $ {d\Gamma_{nnn}^{\rm ring} }/d \Phi$
only in the interval $ 0 \le \Phi \le 120^\circ$ due to the symmetry of the problem.
\label{fig3HdGdphi}}
\end{figure}

The integrated spectra in polar coordinates for the  $\pi^- + {^3{\rm H}} \rightarrow n + n + n $
absorption rates $ {d\Gamma_{nnn}}/dr$ and $ {d\Gamma_{nnn}^{\rm ring} }/d \Phi$ are 
given in Figs.~\ref{fig3HdGdr} and~\ref{fig3HdGdphi}, respectively. Qualitatively, the picture resembles
those for $\pi^- + {^3{\rm He}} \rightarrow p + n + n $ process: the single-nucleon absorption operator based predictions
lie much below the other ones, the PWIAS results are above those taking rescattering among the three outgoing nucleons into account, and
3NF plays no significant role. Both figures highlight a dominant role of the QFS(nn) configuration. 

Finally, our results for the $\Gamma_{nnn}$ rates
are compared in Table~\ref{tableclosure3H} with the corresponding predictions
$\Gamma_{^3{\rm H}}^{\rm closure}$
obtained using the closure approach.
The difference between these two results is very small and does not exceed 0.2\%. The agreement is even more evident than for the total pion absorption rate in $^3$He.  The reason for the better agreement in the $^3$H case might be that, contrary to $^3$He, here we deal with only one, three-body breakup channel. The results confirm the consistence between the three-nucleon bound and scattering states calculated without and with the inclusion of the 3N potential.

\begin{table}
\caption{Absorption rate
	$\Gamma_{nnn}$
	compared to the result $\Gamma_{^3{\rm H}}^{\rm closure}$, 
	obtained using the closure method described in Ref.~\cite{PRC98.054001}
for negative pion absorption in $^3$H calculated with 
the chiral semilocal momentum-space regularized two-nucleon 
	potentials~\cite{SMS} augmented by the consistently regularized 
	three-nucleon force at N$^2$LO~\cite{Maris2021}.
	The initial and final states are computed with the
	two- and three-nucleon forces and the complete transition operator 
	is employed (Full--(SN+2N)--(2NF+3NF)).}
\begin{tabular}{|c|c|c|}
\hline\hline
	       & \multicolumn{2}{l|}{Absorption rates in 10$^{15}$ s$^{-1}$} \\ \cline{2-3}
$\Lambda$ (MeV) & $\Gamma_{nnn}$ & $\Gamma_{^3{\rm H}}^{\rm closure}$ \\ 
\hline
 400   &  1.375   & 1.373 \\ 
 450   &  1.110   & 1.109 \\ 
 500   &  1.002   & 1.001 \\ 
 550   &  1.061   & 1.059 \\
\hline\hline
\end{tabular}
\label{tableclosure3H}
\end{table}

\section{Summary and Outlook}
\label{section5}

We investigated the
$\pi^- + {^2{\rm H}} \rightarrow n + n$,
$\pi^- + {^3{\rm H}} \rightarrow n + n + n$,
$\pi^- + {^3{\rm He}} \rightarrow n + d$
and
$\pi^- + {^3{\rm He}} \rightarrow p + n + n$
capture reactions from the lowest atomic orbitals
under full inclusion of final state interactions.  In the calculations
we employed the LO single-nucleon 
and two-nucleon 
transition operators \cite{Baru:2013zpa} derived  using momentum counting scheme within chiral effective field theory. The nuclear states
were obtained with the chiral semilocal momentum-space regularized
two-nucleon forces up to {N$^4$LO$^+$}~\cite{SMS}, which in the three-nucleon cases
were augmented by the N$^2$LO three-nucleon potentials.
Our calculations have thus rather a ``hybrid'' character, since
the chiral expansion of the nuclear potentials  ignores
the appearance of the intermediate momentum scale $\sim \sqrt{M_\pi M}$
relevant for the pion absorption processes.
Despite this fact our calculations bring important results which should be
confronted with predictions achieved within a more consistent
framework and with experimental data in the future. In particular our results emphasize the decisive role
of the two-nucleon absorption mechanisms in all the studied processes.
Final state interactions effects are also important. Not only do they reduce
the values of the total absorption rates but they alter the shapes of the
differential rates. On the other hand, three-nucleon force effects are
relatively small.

Our LO-MCS result for the rate of the $\pi^- + {^2{\rm H}} \rightarrow n + n$ process evaluated for the N$^4$LO$^+$ wave functions 
is $\Gamma_{nn}  =   (1.3^{+0.2}_{-0.1}\pm 1.0)\, 10^{15} {\rm s^{-1}}$, where the central value  and 
the first  (asymmetric) uncertainty correspond to the average and the spread of the results obtained for four different cutoffs, respectively, 
while the second error is related  to the  LO-MCS truncation uncertainty  estimated very conservatively. 
The predicted values for $\Gamma_{nn}$ show very good agreement with the experimental data from  the hadronic ground-state broadening in pionic 
	deuterium~\cite{Strauch:2010rm,Strauch:2010vu} as well as with the previous EFT calculations \cite{Lensky:2005jc,Baru:2016kru}. 
Using the same chiral NN interactions and pion transition operators we  predict 
	 the pion capture rates on $^3{\rm He}$ and  $^3{\rm H}$, see Eqs.~\eqref{Gam_nd_our}-\eqref{Gam_nd_pnn_our} and \eqref{Gam_nnn_our}. 
While the central values of the predicted capture rates on  $^3{\rm He}$  are found to be systematically smaller than the experimental data, 
our LO-MCS results are consistent with the data within error bars.
The comparison of our normalized predictions for $\Gamma_{pnn}$ with the experimental data from	
Ref.~\cite{Gotta1995}, where relative contributions to the total rate $\Gamma_{pnn}$ from 
four regions in the phase-space were reported, reveals a   rough agreement with the data.

Our predictions for the total absorption rates depend on the order of the two-nucleon potential and on the value of the cutoff parameter $\Lambda$ 
used to construct the NN wave functions. 
The cutoff dependence could be expected since only LO-MCS absorption operators are employed in the present calculations. It would therefore be  important 
to improve the calculations by including the production operators up-to-and-including N$^2$LO-MCS, where two unknown 
NN$\rightarrow\textrm{NN}\pi$ contact interactions start to contribute. 
Once these  contact terms are fixed to low-energy data in $p +p\to  {^2{\rm H}} + \pi^-$ and $p +p\to  p+p  + \pi^0$, respectively, the cutoff dependence 
of the  predicted pion absorption rates on ${^3{\rm He}}$ and ${^3{\rm H}}$ is expected to be  reduced.  Such a study would therefore test the field-theoretical consistency 
 of the proposed EFT framework. It would also allow to  reduce the truncation error of the results for the rates based on the MCS expansion roughly by a factor of 6
 and thus  provide an important test of our understanding of pion production in few-nucleon systems.  

We definitely think that the above mentioned processes are worth further theoretical and experimental studies, since they could bring interesting insights into neutron-neutron and three-neutron interactions. 
Also the corresponding pion absorption on $^4$He, where some experimental data are available \cite{Daum:1995au}, is very interesting because of three different reaction channels.

\acknowledgments
This work was supported in part by BMBF (Grant No.~05P21PCFP4), by DFG and NSFC through funds provided to the
Sino-German CRC 110 ``Symmetries and the Emergence of Structure in QCD" (NSFC
Grant No.~12070131001, Project-ID 196253076 - TRR 110), and by ERC Nuclear Theory (grant No. 885150). 
One of the authors (J.G.) gratefully acknowledges the financial support of the JSPS International Fellowships for Research in Japan (ID=S19149).
The numerical calculations were partly performed on the supercomputers of the JSC, J\"ulich, Germany.

\end{document}